\documentclass[superscriptaddress,twocolumn,notitlepage,amsmath,amssymb,floatfix,10pt,aps,pre]{revtex4-1}

\newcommand{\bvec}[1]{{\mathbf{\string#1} }}
\newcommand{\upd}{\mathrm{d}}
\usepackage{graphicx}
\usepackage{amssymb,amsfonts,amsmath,mathrsfs}
\usepackage{color}
\usepackage{ulem}

\usepackage{relsize}
\usepackage{blindtext}

\DeclareSymbolFont{matha}{OML}{txmi}{m}{it}
\DeclareMathSymbol{\varv}{\mathord}{matha}{118}
\newcommand{\Vext}{V_\text{ext}}

\newcommand{\CPF}{Z}
\newcommand{\LL}{\text{L}}\newcommand{\RR}{\text{R}}\newcommand{\TT}{\text{T}}\newcommand{\HH}{\text{H}}

\begin{document}

\title{Order-preserving dynamics in one dimension - single-file diffusion and caging from the perspective of dynamical density functional theory
}
\author{Ren\'e Wittmann}
\email{rene.wittmann@hhu.de}
\affiliation{Institut f\"ur Theoretische Physik II: Weiche Materie, Heinrich-Heine-Universit\"at D\"usseldorf, D-40225 D\"usseldorf, Germany}
 \author{Hartmut L\"owen}
\affiliation{Institut f\"ur Theoretische Physik II: Weiche Materie, Heinrich-Heine-Universit\"at D\"usseldorf, D-40225 D\"usseldorf, Germany}
\author{Joseph M.\ Brader}\affiliation{Department of Physics, University of Fribourg, CH-1700 Fribourg, Switzerland}
\date{\today}

\begin{abstract}

Dynamical density functional theory (DDFT) is a powerful variational framework to study the nonequilibrium properties of colloids by only considering a time-dependent one-body number density.
 Despite the large number of recent successes, properly modeling the long-time dynamics in interacting systems within DDFT remains a notoriously difficult problem, since structural information, accounting for temporary or permanent particle cages, gets lost.
Here we address such a caging scenario by reducing it to a clean one-dimensional problem, where the particles are naturally ordered (arranged on a line) by perfect cages created by their two next neighbors.
In particular, we construct a DDFT approximation based on an equilibrium system with an asymmetric pair potential, such that the corresponding one-body densities still carry the footprint of particle order.
Applied to a system of confined hard rods, this order-preserving dynamics (OPD) yields exact results at the system boundaries, in addition to the imprinted correct long-time behavior of density profiles representing individual particles.
In an open system, our approach correctly reproduces the reduced long-time diffusion coefficient and subdiffusion, characteristic for a single-file setup.
These observations cannot be made using current forms of DDFT without particle order. 

\end{abstract}

\maketitle

\section{Introduction}

 Ever since Einstein's seminal work on Brownian motion \cite{Einstein}, the dynamics of colloidal particles have been a topic of broad interest \cite{specialchaosissue,frey2005}.
The diffusion of an individual colloid is by now theoretically well understood and many analytical results have been obtained under various external conditions including
(random) external potentials \cite{Zwanzig,Egelhaaf},
flow fields~\cite{Taylor1,Taylor2,Ven,Zimmermann,foister1980},
magnetic fields~\cite{Bonitz,sharmaACT, sharmaRESET, sharmaTHERMO},
 different thermostats for each spatial coordinate~\cite{dotsenko2013,sharmaTHERMO},
 temperature gradients~\cite{Dhont} or
self-motility~\cite{tenHagen2011, vanTeeffelen2008, szamel2014, sprenger2020, sharmaACT, nguyenAOUPS,lorenzoAOUPS}.
However, accurately characterizing the correlated dynamics of many interacting particles is, in general, a highly nontrivial and multilateral theoretical problem. 
One central aspect is the slowing down of the mean-square displacement (MSD) of individual particles as a consequence of the mutual volume exclusion \cite{dhontCOLLOIDS,batchelor1976,hanna1982,elmasri2010,thorneywork2015,mandal2019}.
This results in the reduction of the long-time self-diffusion coefficient accompanied by a subdiffusive regime at intermediate time scales.
 Penultimately, at the glass transition density, it comes to a critical dynamical arrest,  
breaking the ergodicity of the system: a particle cannot explore the full available (phase) space as it is caged by its neighbors,
such that averaging over time does not reflect a full statistical ensemble average.
 This has some severe consequences for theoretical descriptions relying on the ergodicity assumption.

The excluded-volume interaction between the hard cores of overdamped diffusing particles can be incorporated into the governing Fokker-Planck equation in two ways:
either implicitly through reflective boundary conditions \cite{roedenbeckhahnSFD1998,kumar2008,lizina2008,lizina2009,bruna2012a,bruna2012b,bruna2014}
or through an explicit interaction force arising in this special case
from a discontinuous pair-wise interaction potential with values zero and infinity \cite{hansen_mcdonald1986}.
While both strategies are formally exact, their physical interpretation is fundamentally different.
The boundary conditions in the first case mimic the actual collision events between otherwise freely diffusing particles, 
directly resembling the underlying Brownian dynamics. 
The second strategy can be considered the nonequilibrium generalization of (canonical) statistical mechanics, 
as it gives rise to exactly the same definition of the joint probability distribution of the $N$ particle positions in the equilibrium limit.
As such, it applies to any type of pair interaction and can be embedded in more versatile theoretical frameworks 
within which controlled approximation schemes can be developed.
Most notably, the variational method of density functional theory (DFT) \cite{evans79},
which is a cornerstone of modern liquid-state theory \cite{hansen_mcdonald1986},
can be exploited to efficiently describe nonequilibrium dynamics in terms of an ensemble-averaged one-body number density within
dynamical density functional theory (DDFT) \cite{marconi_tarazona,archer_evans,REVIEWDDFT}.

Originally derived to describe overdamped Brownian systems of spherical particles, 
the diversity of DDFT was increased through various extensions towards, e.g., Newtonian fluids \cite{archer2006newtonian}, 
anisotropic particles \cite{rexloewen_anisoDDFT,wittkowskiloewen_activeanisoDDFT}, hydrodynamic interactions \cite{rexloewenDDFT,goddard2012}.
Thereby, this framework has become an important tool to better understand spatiotemporal aspects of a broad range of phenomena \cite{REVIEWDDFT}, including
quasicrystals \cite{archerknobloch2015},
swimming organisms \cite{menzel2016,hoell2017},
cellular dynamics \cite{alsaedi2018},
and epidemic spreading \cite{vrugt2020}.
Owing to further developments over the last years, 
several intrinsic deficiencies of DDFT have been overcome.
First, more and more sophisticated equilibrium functionals became available, particularly for hard interactions \cite{roth10,roth12,wittmann16,wittmann2017}, which is a basic requirement for accurate dynamics. 
Second, while the time-dependent density profile provided by DDFT represents collective motion,
individual transport properties become accessible in the dynamical generalization  \cite{archerDTP2007,hopkinsDTP2010,reinhardtbrader2012,stopper2015PRE,stopper2015JCP,stopper_bulk} of Percus' test particle theory \cite{percusTP}.
Third, despite the inherently grand-canonical nature of DFT \cite{canonical1,canonical2} canonical information is available through an inversion method \cite{delasheras2014},
which can also be exploited in the context of DDFT \cite{delasherasbrader2016,schindlerproject}.
This particle-conserving dynamics (PCD) is a crucial step towards a realistic description of the Brownian {reference} system.
Finally, the way in which the interaction force in DDFT is constructed from an equilibrium free-energy functional
implies that nonequilibrium correlations are replaced by equilibrium ones, which is called an adiabatic approximation.
To include the missing superadiabatic forces \cite{PFTxPRL,PFTxPRLseparation},
the formally exact framework of power functional theory has been developed, which contains (adiabatic) DDFT as a limiting case \cite{PFT1,PFTtestparticle}.
This generalized variational approach also includes the one-body current but 
still requires a workable free-energy term in addition to the recently introduced approximations for the nonequilibrium corrections \cite{pftapp1,pftapp2,pftapp3}. 
Therefore, power functional theory also relies on the basic suitability of the underlying DDFT for the problem of interest.
What remains to be better understood is the failure of DDFT to intrinsically describe the slow-down of the long-time self diffusion in general \cite{hopkinsDTP2010,stopper2015PRE,stopper2015JCP,stopper_bulk} 
and the difficulty to model glassy states by DFT methods in particular \cite{singh1985,archerDTP2007,hopkinsDTP2010,REVIEWDDFT}.

\begin{figure}[t]
\centering
\includegraphics[width=0.45\textwidth] {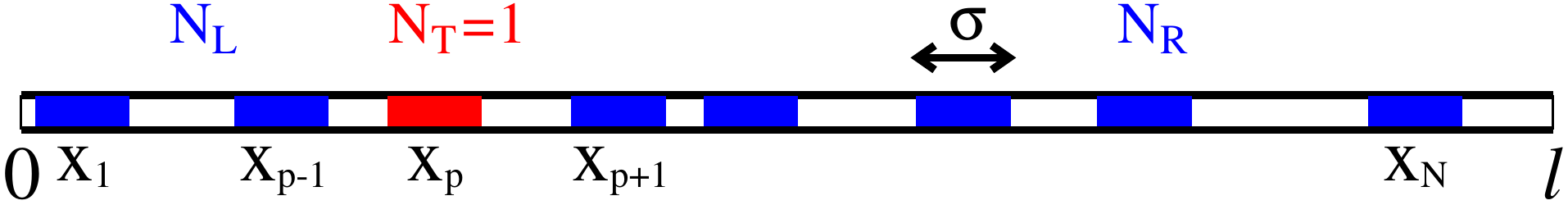}
\caption{ Sketch of the ordered one-dimensional system considered here. The $N\!=\!N_{\LL}+1+N_{\RR}$ hard rods of length $\sigma$ distributed on a line of length $l$ are separated into three species $\nu$ whose order is preserved. 
This means that the particle labeled $p$ is considered as an individual \textit{tagged} species $\nu\!=\!{\TT}$ with $N_{\TT}\!=\!1$, which is always located in between the particles of the species $\nu\!=\!{\LL}$ on the left and $\nu\!=\!{\RR}$ on the right. \label{fig_SYSTEM}
}
\end{figure}

Sometimes perceived only as an oversimplified playground, one-dimensional systems possess some outstanding dynamical properties in their own right, 
 emerging from their characteristic particle order (understood in this context as fixed relative positions of particles on a line, cf.\ Fig.~\ref{fig_SYSTEM}).
Most prominently, the resulting impossibility of particles confined to a narrow channel to overtake gives rise 
to unique transport properties emerging in a broad range of systems \cite{hodgkin1955,hahn1996,kukla1996,wei2000,das2010,delfau2010,gupta2013,lutz2014}.
 For overdamped Brownian systems, in particular, this so-called single-file diffusion (SFD) is characterized by the universal subdiffusive exponent $1/2$ of the MSD \cite{harris1965,percusSFD,alexander1978,kollmann2003}. 
In this setup, the exact tagged-particle dynamics for hard interactions can be determined analytically \cite{roedenbeckhahnSFD1998,kumar2008,lizina2008,lizina2009,kumar2008,percus2010,krapivsky2014,krapivsky2015}.
More complex SFD scenarios with, e.g., different diffusivities \cite{ambjornsson2008a}, finite-ranged interactions \cite{ambjornsson2008b} or an external drive \cite{ryabov2011},
are exactly solvable on the two-body level.
 In addition, systems with external potentials \cite{taloni2006}, soft interactions \cite{nelissen2007} 
or a tagged particle with significantly larger mass \cite{foulaadvand2013} have been investigated by means of computer simulations.
A description of SFD by means of DDFT, however, exposes some surprisingly deep problems.
Despite being based on the exact Percus functional \cite{percus,percus2} in the generalized version for mixtures \cite{percusmix}, 
the PCD of Ref.~\cite{schindlerproject} only appropriately describes the dynamics of individual particles for short times,
which means that the characteristic subdiffusive long-time behavior of SFD is not accessible from this theory.
This behavior cannot be attributed to the adiabatic assumption {alone}, 
which is best understood by considering the SFD of an (ordered) ideal gas \cite{roedenbeckhahnSFD1998,kumar2008} (see also Fig.~\ref{fig_MSDsum} below).
In this case, PCD {just} predicts the ordinary diffusion of an ideal gas, 
 which is in fact the generic long-time limit for self diffusion in DDFT \cite{hopkinsDTP2010,reinhardtbrader2012,stopper2015PRE},
if no manual corrections are performed on the diffusion coefficient \cite{stopper2015JCP,stopper_bulk}.
Instead, the erroneous long-time behavior from PCD (and DDFT) can be explained \cite{schindlerproject,reinhardtbrader2012} by the fundamental assumptions of the
underlying statistical mechanics in the equilibrium limit that the system is ergodic and mixing.
In other words, the DFT with the Percus functional reproduces exactly all statistical mechanical results in one dimension,
whereas any dynamical theory based on pair potentials and ensemble averages cannot produce exact 
behavior of particles that are distinguishable by their order. 
A similar problem occurs \cite{cremer2017} (and persists in higher dimensions \cite{goh2019}) when mapping the elastic energy in bead-spring models for ferrogels onto a pair potential of indistinguishable particles for a DFT implementation.

Here, we describe how SFD can be addressed from the perspective of DDFT.
To this end we introduce an asymmetric interaction potential to keep {the different particles separated.
This strategy} breaks the ergodicity at the heart of statistical mechanics {in a generic way,
without manually reducing each particle's phase space by an \textit{a priori} adaption of the configurational integrals.}
Then we develop a variational framework to describe ordered equilibrium systems by combining (canonical) DFT for a conditional one-body density with a subsequent {computation of the remaining} configurational integral {representing a tagged particle}. 
Employing this strategy together with the adiabatic approximation allows us to determine the order-preserving dynamics (OPD) of interacting particles in a narrow channel 
and interpret the resulting time-dependent density profiles and mean-square displacement. 
Despite the availability of exact results for the problem considered, our goal here is to learn more about the fundamentals of DDFT. 
We are particularly concerned with the possibility to reproduce caging effects in a variational framework, 
and {with assessing} the reliability of the adiabatic approximation,
in order to provide a solid basis for employing the developed approach within the more general power functional theory.

The paper is arranged as follows.
In Sec.~\ref{sec_statmech} we lay the static foundations
{and illustrate how the order of individual particles in equilibrium can be formally accounted for in the language of statistical mechanics.
The busy reader may skip this section at first glance and later get back to the referenced content at will.
Then we use the obtained} ordered distributions in Sec.~\ref{sec_OPD} to construct a dynamical theory with both conserved particle number and preserved order,
and study the time evolution of the density profiles of confined hard rods and the MSD in open systems.
{Some figures showing additional data are attached to this manuscript \cite{SI}.}
We conclude in Sec.~\ref{sec_conclusion} on the implications of our results on addressing the caging scenario by variational theories.
Although the mathematical background of DFT is not explicitly required to understand these calculations,
 our results are thoroughly interpreted in this context.
The reader unfamiliar with DFT methods is thus referred to appendix~\ref{app_DDFT}.

\section{Ordered one-body densities in statistical mechanics \label{sec_statmech}} 

\subsection{Conditional probabilities in the canonical ensemble \label{sec_c0}}

Before dealing with the problem of particle order and distinguishability in statistical mechanics, we 
recapitulate and extend an insightful classical interpretation of Percus' test-particle limit \cite{percusTP} by Henderson \cite{henderson1983},
which helps to put our later expressions into a broader context, particularly regarding the role played by (D)DFT.
Consider in any spatial dimension $d$ a bulk fluid of $N$ indistinguishable particles in a volume $V$ and at temperature $T$,
which interact with a pairwise-additive potential
\begin{equation}
 \mathcal{U}_N(\bvec{r}^N)=
 \sum_{\substack{i,j=1
 \\ i<j}}^{N
 } u
 \left(|\bvec{r}_{i}-\bvec{r}_{j}|\right)
 \label{eq_UNr}
\end{equation}
with the isotropic pair potential $u(r)$
and are subject to a total external potential 
\begin{equation}
 \mathcal{V}_N(\bvec{r}^N)=\sum_{i=1}^{N}\Vext^{(i)}\left(\bvec{r}_{i}\right)
 \label{eq_VNr}
\end{equation}
where each particle may by affected by a different one-body field $\Vext^{(i)}(\bvec{r})$.

Let us define a \textit{conditional} canonical partition function $\mathcal{\CPF}_{N-1}(T,V,N|\bvec{r}_0)$
for a system of $N$ physically identical particles one of which (labeled $p\in\{1,\ldots,N\}$) is pinned at some position $\bvec{r}_0$
according to
\begin{align} 
 e^{-\beta\Vext^{(p)}(\bvec{r}_0)}\mathcal{\CPF}_{N-1}(\bvec{r}_0)=\frac{\int\mathrm{d}\bvec{r}^{N}\,e^{-\beta\mathcal{U}_{N}-\beta\mathcal{V}_{N}}\,\delta(\bvec{r}_p-\bvec{r}_0)}{(N-1)!\,\Lambda^{(N-1)d}}\,,
\label{eq_config1r}
\end{align}
where $\Lambda$ denotes the thermal wavelength \cite{hansen_mcdonald1986}.
 Here and in the following, we omit the thermodynamic variables in the argument of partition functions, when these are clear from the context.
The ordinary canonical partition function $\CPF_{N}(T,V,N)$ of the $N$ freely moving particles then follows from
{integrating over the remaining configurational variable in Eq.~\eqref{eq_config1r} according to}
\begin{align}
\CPF_N=\int\mathrm{d}\bvec{r}_0\,\frac{\mathcal{\CPF}_{N-1}(\bvec{r}_0)}{\Lambda^d}\,e^{-\beta \Vext^{(p)}(\bvec{r}_0)}\,.
\label{eq_QNr}
\end{align}
In this system, the particle at $\bvec{r}_0$ (with an arbitrary label $p$) can be interpreted as a tagged particle, for which we introduce the superscript $(\TT)$
and identify $\Vext^{(\TT)}\equiv\Vext^{(p)}$.
The one-body density $\rho_N^{(\TT)}(\bvec{r})$ of the tagged particle follows from the average of the density operator $\hat{\rho}^{(\TT)}
=\delta(\bvec{r}-\bvec{r}_0)$ as
\begin{align}\label{eq_rhoWcanr}
 \rho_N^{(\TT)}(\bvec{r})=\frac{\mathcal{\CPF}_{N-1}(\bvec{r})}{\Lambda \CPF_N}\,e^{-\beta \Vext^{(\TT)}(\bvec{r})}\,.
\end{align}
Now consider the one-body density $\rho_N^{(\HH)}(\bvec{r})$ of the remaining $N-1$ host particles $(\HH)$ as the average of $\hat{\rho}^{(\HH)}
=\sum_{i\neq p}\delta(\bvec{r}-\bvec{r}_i)$.
The result can be rearranged to the instructive form
\begin{align}
 \rho_N^{(\HH)}(\bvec{r})=\int\mathrm{d}\bvec{r}_0\,\rho_N^{(\TT)}(\bvec{r}_0)\,\varrho_N^{(\HH)}(\bvec{r}|\bvec{r}_0)\,,
 \label{eq_rhoIcanGENr}
\end{align}
where $\varrho_N^{(\HH)}(\bvec{r}|\bvec{r}_0)$ is the conditional one-body density of the $N-1$ particles given the tagged particle is located at $\bvec{r}=\bvec{r}_0$.
In the ensemble determined by Eq.~\eqref{eq_config1r}, this quantity can be defined as
\begin{align} 
 \varrho_N^{(\HH)}(\bvec{r}|\bvec{r}_0)=\frac{\int\mathrm{d}\bvec{r}^{N}\,e^{-\beta\mathcal{U}_{N}-\beta\mathcal{V}_{N}}\,\delta(\bvec{r}_p-\bvec{r}_0)\,\hat{\rho}^{(\HH)}(\bvec{r},\bvec{r}^{N-1})}{(N-1)!\,\Lambda^{(N-1)d}\,e^{-\beta\Vext^{(p)}(\bvec{r}_0)}\mathcal{\CPF}_{N-1}(\bvec{r}_0)}\,.
\label{eq_varrhoiCr}
\end{align}
The analogy with probability theory becomes obvious when identifying the expression under the integral in Eq.~\eqref{eq_rhoIcanGENr} as the two-body density
\begin{align}
 \frac{1}{N}\,\rho_N^{(2)}(\bvec{r},\bvec{r}_0)=\rho_N^{(\TT)}(\bvec{r}_0)\,\varrho_N^{(\HH)}(\bvec{r}|\bvec{r}_0)\,,
 \label{eq_rhoIcanGENr2}
\end{align}
 which is the joint probability to find the tagged particle at $\bvec{r}_0$ and {a host particle} at $\bvec{r}$.
 Note that the densities considered here are not probability densities, i.e., their normalization is related to the particle numbers of the species considered \cite{hansen_mcdonald1986}.
 The factor $N$ in Eq.~\eqref{eq_rhoIcanGENr2} is required since we made a particular choice for the tagged particle.
  There also exist related grand-canonical expressions, as further discussed in Sec.~\ref{sec_gc}.

What we learn from this exercise, in particular {regarding} (D)DFT, are the following four points.
First, Eqs.~\eqref{eq_rhoWcanr} and \eqref{eq_rhoIcanGENr} provide an indirect route to calculate the density of a fluid by means of auxiliary conditional quantities,
which can be calculated in a theory assuming that a pinned particle acts as an external potential, while the remaining statistical integration is carried out by hand.
This conditional DFT is described in appendix~\ref{app_DDFT}.
Following the argumentation in Ref.~\cite{archerevans2017}, such an approach could improve the density profiles predicted by an approximate functional.
Second, Eq.~\eqref{eq_rhoIcanGENr2} constitutes a generalization of Percus' test-particle approach, which is recovered for equal external potentials $\Vext^{(1)}=\Vext^{(2)}=\ldots=\Vext^{(N)}$.
This can be easily shown for a homogeneous bulk fluid with constant density $\rho_\text{b}=N/V$ and isotropic radial distribution function $g(r)$.
Setting $\rho_N^{(2)}=\rho_\text{b}^2g(r)$ and $\rho_N^{(\TT)}=\rho_\text{b}/N$ yields the famous result~\cite{percusTP}
\begin{align}
\rho_\text{b}\, g(r)=\varrho_N^{(\HH)}(\bvec{r}|\bvec{r}_0)\,,
 \label{eq_rhoIcanGENr2TPT}
\end{align}
where the conditional density, Eq.~\eqref{eq_varrhoiCr}, on the right-hand side is again interpreted as an inhomogeneous density in the test system. 
Third, generalizing this idea to a dynamical test particle theory allows to explicitly treat $\rho_N^{(\TT)}(\bvec{r}_0)$ as a canonical quantity, circumventing the problem of self-interaction in DDFT \cite{reinhardtbrader2012,stopper2015PRE,stopper2015JCP,stopper_bulk}. 
Finally, restricting the range of the tagged-particle distribution through the interaction with the host particles
allows to model localization effects.
This strategy is particularly insightful in one spatial dimension and serves as the basis for the calculations in remainder of this work.

\begin{figure}[t]
\centering
\includegraphics[width=0.45\textwidth] {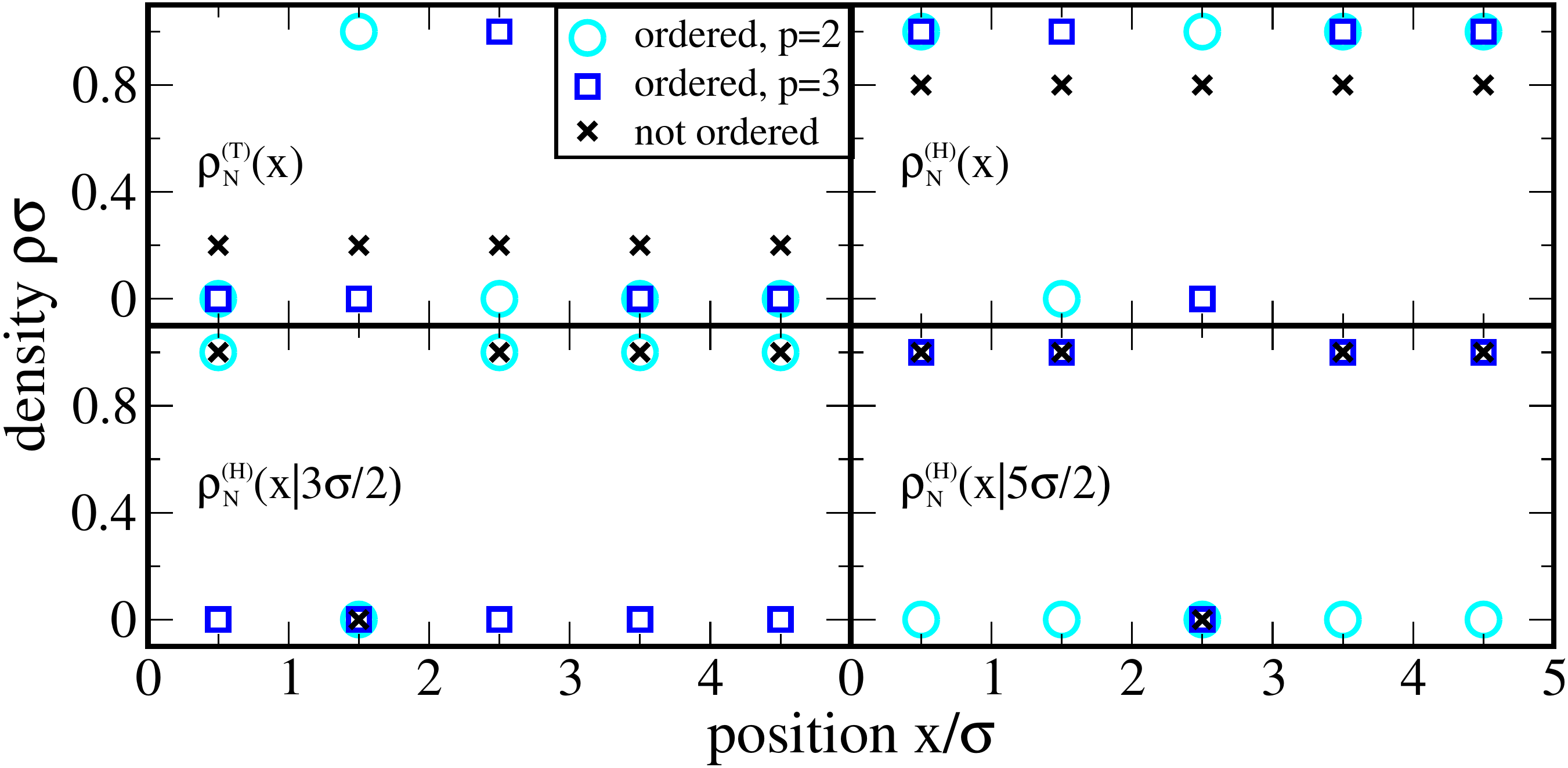}
\caption{  Schematic illustration of the introduced (conditional) one-body densities 
$\rho_N^{(\TT)}(x)$ (top left), $\rho_N^{(\HH)}(x)\!\simeq\!\rho_N^{(\LL)}(x)+\rho_N^{(\RR)}(x)$ (top right) and $\varrho_N^{(\HH)}(x|x_0)$ (bottom left for $x_0\!=\!3\sigma/2$ and right for $x_0\!=\!5\sigma/2$) in one dimension.
We compare the ordered profiles for a tagged particle labeled $p\!=\!2$ (circles) and $p\!=\!3$ (squares)
to those without particle order, where $p$ is arbitrary {(crosses)}
considering an idealized packed system of $N\!=\!5$ hard rods of length $\sigma$ confined between two walls of separation $l\!=\!5\sigma$.
The sharp density peaks at $x\!=\!\sigma/2+n\sigma$ with $n\in\{0,1,\ldots,4\}$ are {shown as} single points representing the value of their integral.
 If in OPD the $N_\alpha$ particles do not fit into the respective subsystem, either $\rho_N^{(\LL)}$ or $\rho_N^{(\RR)}$ is undefined and we set $\rho_N^{(\HH)}$ to zero.
 {See text for further details.}
\label{fig_DENSITIES}
}
\end{figure}

\subsection{Ordered ensemble for nonergodic hard-body mixtures \label{sec_c}} 

Now we turn to $d=1$ dimension with the single spatial coordinate $x$ replacing $\bvec{r}$.
In this case, illustrated in Fig.~\ref{fig_SYSTEM}, the $N$ hard rods of length $\sigma$ are ordered on a line of length $l$ (or within an infinitely small channel), such that they cannot overtake.
{For such a system, it has been elaborated in Ref.~\cite{schindlerproject} that 
a simple relabeling (distinction of particles only in terms of combinatorial factors) is not sufficient to determine the proper equilibrium distributions of individual particles. 
To better understand this crucial point, we discuss below in detail how it boils down to the interpretation of Eq.~\eqref{eq_rhoIcanGENr},
focusing on the extreme example of a system that cannot hold more than $N$ particles, cf.~Fig.~\ref{fig_DENSITIES}.

In the case of completely indistinguishable but labeled particles, the} 
conditional density $\varrho_N^{(\HH)}(x|x_0)$ carries implicit information about the particle order, which is lost in the final integration over $x_0$.
If the tagged particle is fixed at $x_0$, the numbers $N_{\LL}$ and $N_{\RR}$ of particles fitting on its left and right are well defined, where $N_{\LL}+N_{\RR}=N-1$.
{However, the identity $\rho_N^{(\HH)}(x)=(N-1)\,\rho_N^{(\TT)}(x)$ holds for the resulting one-body density profiles}, if the external potentials {of each species} are equal.
This loss of microscopic information directly reflects the fact that the underlying statistical ensemble is ergodic and mixing.
{This can be most directly seen in the} symmetry of the
pairwise interaction potential
\begin{equation}
  u(|x_i-x_j|)=\left\{ 
 \begin{array}{cl} 0 & \quad |x_i-x_j|>\sigma \\
      \infty & \quad |x_i-x_j|<\sigma
 \end{array}\right.\,,\label{eq_u}
\end{equation}
which only forbids particle overlaps, but gives an equal statistical weight to {one particular configuration of two particles and the configuration} with both positions interchanged.

{The relation between the one-body densities and the conditional densities is further illustrated in Fig.~\ref{fig_DENSITIES} for the case of $N=5$ 
perfectly localized particles.
Here, the black crosses in the top row indicate the probabilities $1/5$ and $4/5$ that a particle at any given position is the tagged particle or one of the host particles, respectively.
In contrast, the conditional densities in the bottom row explicitly depend on the location of the tagged particle.
 Here, the probability to find a host particle at the exact position of the tagged particle is zero, while it is one at the other possible positions.
}

To imprint the ordered property of the particles into their one-body densities, {there are two mathematically equivalent possibilities, both with the same combinatorial prefactors indicating distinguishable species. The first way amounts to}
  include the particle order explicitly in {the boundaries of} the statistical integrals used to calculate the averages.
 In general, this implies that each particle is {truly} distinguishable from the others by its relative position and thus formally corresponds to an individual species.
 {Alternatively, these restricted integral boundaries can be absorbed into the interaction potential between the different particles \cite{schindlerproject}.
As this second interpretation of the particle order can be directly transferred to the framework of DFT \cite{DFTfootnote},
we will adapt it in the following presentation for the sake of providing a general picture, although the calculations made in this paper do not require such a particular choice. 
In addition,} instead of solving the full $N$-body problem,
we conveniently consider a mixture of three species
with one tagged particle ($N_{\TT}=1$) at $x_0$ confined in the middle of the other species holding $N_{\LL}$ and $N_{\RR}$ particles.
Although our considerations apply to any {type of pair} interaction (as long as some mechanism explicitly forbids overtaking),
we restrict ourselves in the following to hard rods to keep the notation compact.

{We start by splitting} the interaction potential $\mathcal{U}_N(x^N)$ from Eq.~\eqref{eq_UNr}
into two independent parts.
First we consider only a partial interaction potential $\mathcal{U}_{N_{\LL}}+\mathcal{U}_{N_{\RR}}$ between $N-1$ particles
by removing all pair potentials which we associate with the coordinate $x_p$ of the tagged particle (where $p=N_{\LL}+1=N-N_{\RR}$).
Then we introduce a new potential $\mathcal{W}_N(x^N)$ denoting the interaction with the tagged particle, thereby taking the role of a conditional external potential $\mathcal{W}_N(x^{N_{\LL}+N_{\RR}}|x_p)$ acting on the host particles.
Instead of constructing $\mathcal{W}_N$ in terms of the symmetric pair potentials $u(|x|)$, defined in Eq.~\eqref{eq_u},
 which would simply amount to a reinterpretation of the total interaction $\mathcal{U}_N$, we set
\begin{equation}
 \mathcal{W}_{N}(x^N)= \sum_{i=1}^{N_{\LL}} w_{\LL}(x_i-x_p) + \sum_{i=N_{\LL}+2}^{N} w_{\RR}(x_i-x_p)\,.
 \label{eq_WNalpha}
\end{equation}
Thereby, we define the order-preserving pair potentials
\begin{equation}
 w_\alpha(x_i-x_p)=\left\{ 
 \begin{array}{cl} 0 & \quad s_\alpha\,(x_i-x_p)>\sigma \\
      \infty & \quad s_\alpha\,(x_i-x_p)<\sigma
 \end{array}\right.\,,\label{eq_w}
\end{equation}
where $\alpha\in\{\LL,\RR\}$ serves as a specific species label and the sign function $s_\alpha$ in the potential is $s_\LL=-1$ or $s_\RR=+1$.
{With this choice, the} modification 
\begin{equation}
 \mathcal{U}_N\rightarrow\mathcal{U}_{N_{\LL}}+
 \mathcal{U}_{N_{\RR}}+
  \mathcal{W}_{N}
  \end{equation}
of the standard statistical setup gives rise to an \textit{ordered ensemble}, ensuring that the intended configuration is actually recovered in the process of statistical averaging.

In the ordered ensemble for three components of uniform hard rods in $d=1$ dimension specified above, the conditional canonical partition function, defined analogously to Eq.~\eqref{eq_config1r},
can be written in the factorized form
\begin{align}
 \mathcal{\CPF}_{N-1}(x_0)&=\mathcal{\CPF}^{(\LL)}_{N_{\LL}}(x_0)\,\mathcal{\CPF}^{(\RR)}_{N_{\RR}}(x_0)\,,\cr
 \mathcal{\CPF}^{(\LL)}_{N_{\LL}}(x_0)&=\int_{-\infty}^{x_0-{\sigma}}\mathrm{d}x^{N_{\LL}}\,\frac{e^{-\beta\mathcal{U}_{N_{\LL}}-\beta\mathcal{V}_{N_{\LL}}}}{N_{\LL}!\,\Lambda^{N_{\LL}}}\,,\cr
\mathcal{\CPF}^{(\RR)}_{N_{\RR}}(x_0)&= \int^{\infty}_{x_0+\sigma}\mathrm{d}x^{N_{\RR}}\,\frac{e^{-\beta\mathcal{U}_{N_{\RR}}-\beta\mathcal{V}_{N_{\RR}}}}{N_{\RR}!\,\Lambda^{N_{\RR}}} \,,
\label{eq_config}
\end{align}
where the relevant order-preserving potentials
\begin{equation}
 \mathcal{W}^{(\alpha)}_{N_\alpha}({x_p,} x^{N_\alpha}):= \sum_{i=1}^{N_\alpha} w_\alpha(x_i-x_p)
 \label{eq_WNalphaSEP}
\end{equation}
have been absorbed into the respective integral boundaries.
As in Eq.~\eqref{eq_QNr},
the canonical partition function $\CPF_N$ of $N$ indistinguishable particles is recovered upon {carrying out the final configurational} integral.
This {result} is not surprising, since reverting to ordered particles is just a mathematical trick to calculate the $N$ integrals in a one-component system \cite{percus,percus2},
{but it nicely illustrates that the information about order, uncovered on the conditional level, is hidden in the full averaging process.}
We further stress that the canonical partition function for a true mixture of actually indistinguishable (but mixed) particles is different from $\CPF_N$,
which we explicitly evaluate in appendix \ref{app_ideal} for point particles.

Apart from the form of $\mathcal{\CPF}_{N-1}$, the difference between the ordered ensemble 
and an ergodic system
becomes apparent {\cite{DFTfootnote}} on the level of the one-body densities 
\begin{align}
 \rho_N^{(\nu)}(x)&=\int\mathrm{d}x_0\,\frac{\mathcal{\CPF}_{N-1}(x_0)}{\Lambda \CPF_N}\,e^{-\beta \Vext^{(\TT)}(x_0)}\,\varrho_N^{(\nu)}(x|x_0)\,,\label{eq_rhoIcanGEN}
\end{align}
where $\nu\in\{{\LL},{\TT},{\RR}\}$.
The conditional densities in the two subsystems left ($\alpha={\LL}$) and right ($\alpha={\RR}$) of the tagged particle read
\begin{align} 
 \varrho_N^{(\alpha)}(x|x_0)=\frac{\int\mathrm{d}x^{N_\alpha}\,e^{-\beta\mathcal{U}_{N_\alpha}-\beta\mathcal{V}_{N_\alpha}-\beta\mathcal{W}^{(\alpha)}_{N_\alpha}}\,\hat{\rho}^{(\alpha)}(x,x^{N_\alpha})}{N_\alpha!\,\Lambda^{N_\alpha}\,\mathcal{\CPF}^{(\LL)}_{N_{\LL}}(x_0)}\,
\label{eq_varrhoiC}
\end{align}
 and we formally define $\varrho_N^{(\TT)}(x|x_0):=\delta(x-x_0)$.

 In this ordered setup, $\rho_N^{(\TT)}$ obviously depends on the chosen particle index $p$ (and thus on the predetermined numbers $N_{\LL}$ and $N_{\RR}$).
Hence, there is no simple relation
$\rho_N^{(\LL)}+\rho_N^{(\RR)}\neq (N-1)\,\rho_N^{(\TT)}$ between the density profiles as in standard canonical treatment allowing for the intermixing of particles.
 However, the general identity 
 \begin{align}
  \sum_\nu\rho_N^{(\nu)}=\sum_{p=1}^{N} \rho_N^{(\TT)}=\rho_N
  \label{eq_sumrhoc}
 \end{align}
 holds in both cases.

 Returning to {the example system of perfectly localized particles from Fig.~\ref{fig_DENSITIES}}, the interpretation 
 of the conditional densities from Eq.~\eqref{eq_varrhoiC} is the following.
Fixing $x_0$ such that there are $N_{\LL}$ and $N_{\RR}$ particles at the sides of the tagged particle,
{the probability to find the tagged particle with label $p=N_{\LL}+1$ at position $x=x_0$ is one, while it is zero at $x=x_0$ for the host particles (see, e.g., the blue squares in the top row for $p=3$ and $x_0=5\sigma/2$).}
  The {corresponding} conditional partition function and densities {for a tagged particle at $x=x_0$} are the same as if $\mathcal{W}^{(\alpha)}_{N_\alpha}$ would be made of symmetric pair potentials,
   {compare the blue squares to the black crosses on the bottom right}.
However, for other positions, say {$x_0=3\sigma/2$, of the same tagged particle} we have $\mathcal{\CPF}_{N-1}=0$, such that $\rho_N^{(\TT)}$ is zero {at this point}.
Moreover, also {$\varrho_N^{(\LL)}(x|3\sigma/2)$ is then} undefined, as the $N_{\LL}$ particles do not fit into the subsystem, {and we set $\varrho_N^{(\HH)}(x|3\sigma/2)=0$ (blue squares on the bottom left)}.
In contrast, for symmetric pair potentials {(black crosses)} there would be a finite contribution with other numbers of neighbors.
Such a contribution, in turn, could be associated with another tagged particle 
in the ordered ensemble {for a different label $p$ (cyan circles in the bottom row for $p=2$)}.

\subsection{Ordered ensembles with fluctuating particle numbers \label{sec_gc}} 

Having established an ensemble which provides ordered distributions with a fixed particle number,
we ask the question of what is (are) the corresponding grand canonical ensemble(s), in which the particle numbers can fluctuate.
While the canonical partition function $\CPF_N(T,V,N\!=\!N_{\LL}\!+\!N_{\TT}\!+\!N_{\RR})$ of the ordered ensemble is equal to that of indistinguishable particles (only the ensemble averages of nontrivial operators may be different),
it turns out that the corresponding grand partition function
is neither unique (there are different sensible ways to introduce such a quantity) nor equivalent to the grand partition function 
\begin{align}\label{eq_XiGC}
 \Xi(T,V,\mu)=\sum_{N=0}^\infty e^{\beta\mu N}\,\CPF_N
\end{align}
 of a single component with the chemical potential~$\mu$. 
To see this, we will first define {below} the two possible partition functions {$\Xi_\text{gcg}(T,V,N_{\TT}\!=\!1,\mu_{\LL},\mu_{\RR})$} and $\Xi_\text{ggg}(T,V,\mu_{\LL},\mu_{\TT},\mu_{\RR})$ corresponding to our ordered three-component mixture,
assuming that the particle number fluctuates in two and three species, respectively (as indicated by the subscripts $``\text{c}"$ for canonical and $``\text{g}"$ for grand canonical treatment of a species).
A third possibility, {$\Xi_\text{cgc}(T,V,N_{\LL},N_{\RR},\mu_{\TT})$}, with one fluctuating species could also be considered but does not turn out to be useful in the present context.
Recall that up to this point the particle number $N_{\TT}$ of the species holding the tagged particle has been fixed as $N_{\TT}\equiv1$.

Now let us properly define 
the ordered ensembles of interest by calculating
\begin{align}\label{eq_Xigcg}
\Xi_\text{gcg}&=\sum_{N_{\LL}=0}^\infty\sum_{N_{\RR}=0}^\infty e^{\beta\mu_{\LL} N_{\LL}}\, e^{\beta\mu_{\RR} N_{\RR}}\,\CPF_N\,,\\
\label{eq_Xiggg}
\Xi_\text{ggg}&=\sum_{N_{\LL}=0}^\infty\sum_{N_{\TT}=0}^\infty \sum_{N_{\RR}=0}^\infty e^{\beta\mu_{\LL} N_{\LL}}\, e^{\beta\mu_{\TT} N_{\TT}}\, e^{\beta\mu_{\RR} N_{\RR}}\,\CPF_N\,.
\end{align}
The common starting point $\CPF_N$, which differs from the partition function $\prod_\nu \CPF_{N_\nu}$ of a true mixture,
 reflects that we are dealing with an ordered system. 
Although the sequence of the three ordered components (${\LL}$ on the left, ${\TT}$ in the middle and ${\RR}$ on the right) is not apparent from these functions, it is implied at this point by the label of the chemical potentials.
In appendix \ref{app_ideal}, we explicitly show for ordered point particles that, even if all chemical potentials are equal, 
$\Xi(\mu)\neq\Xi_1(\mu,\mu)\neq\Xi(\mu,\mu,\mu)$
are indeed different functions.

The explicit particle order in the fluctuating ensembles is reflected by the one-body densities,
which can be obtained from the ordered canonical ones in Eq.~\eqref{eq_rhoIcanGEN}
   according to
\begin{align}\label{eq_rhoGCG}
\rho_\text{gcg}^{(\nu)}(x)&=\sum_{N_{\LL}=0}^\infty\sum_{N_{\RR}=0}^\infty p_\text{gcg}\,\rho_N^{(\nu)}(x)\,,\\
\rho_\text{ggg}^{(\nu)}(x)&=\sum_{N_{\LL}=0}^\infty\sum_{N_{\TT}=0}^\infty\sum_{N_{\RR}=0}^\infty p_\text{ggg}\,\rho_N^{(\nu)}(x)\,, \label{eq_rhoGGG}
\end{align}
where 
\begin{align}
 p_\text{gcg}(\{N_\alpha\},\{\mu_\alpha\})&=e^{\beta\mu_{\LL}N_{\LL}}\,e^{\beta\mu_{\RR}N_{\RR}}\frac{\CPF_N}{ \Xi_\text{gcg}} \label{eq_pN1N2}\,,\\
 p_\text{ggg}(\{N_\nu\},\{\mu_\nu\})&=e^{\beta\mu_{\LL} N_{\LL}}\, e^{\beta\mu_{\TT} N_{\TT}}\, e^{\beta\mu_{\RR} N_{\RR}}\frac{\CPF_N}{\Xi_\text{ggg}} \label{eq_pN1N2N3}
\end{align}
are the probabilities to find $\{N_\alpha\}$ and $\{N_\nu\}$ particles at given chemical potentials $\{\mu_\alpha\}$ or $\{\mu_\nu\}$, respectively, recalling that $\alpha\in\{{\LL},{\RR}\}$ and $\nu\in\{{\LL},{\TT},{\RR}\}$.
 We stress that for ordered point particles in a box, we obtain in appendix~\ref{app_ideal} closed expressions for $\rho_\text{gcg}^{(\alpha)}$ and $\rho_\text{ggg}^{(\alpha)}$,
while the canonical counterpart $\rho_N^{(\alpha)}$ can only be cast in a series representation {similar to} Eq.~\eqref{eq_sumrhoc}.

 While the ordered canonical (in the present notation the $``\text{ccc}"$) ensemble provides the exact equilibrium limit for dynamics with conserved particle numbers,
this grand-canonical generalization constitutes a crucial step towards the description of (semi-) infinite systems
and provides the connection to the framework of DFT. 
 In detail, treating the tagged-particle problem in the gcg ensemble appears to be a convenient
 approach for systems with large particle numbers, 
 particularly, when we are also interested in the distributions of the host particles.
 Moreover, it bridges the gap between the proposed treatment and a conditional DFT, which outputs grand-canonical conditional densities $\varrho^{(\alpha)}(x|x_0)$ and partition functions $\varXi^{(\alpha)}(x_0)$, as detailed in appendix~\ref{app_DDFT}.
 To rewrite Eqs.~\eqref{eq_Xigcg} and~\eqref{eq_rhoGCG} in terms of these quantities, we can proceed analogously to the canonical formalism in Secs.~\ref{sec_c0} and~\ref{sec_c}, which yields
  \begin{align}
 \Xi_\text{gcg}&=\int\mathrm{d}x_0\,\frac{\varXi^{(\LL)}(x_0)\,\varXi^{(\RR)}(x_0)}{\Lambda}\,e^{-\beta \Vext^{(\TT)}(x_0)}
 \label{eq_gconfig}
\end{align}
and
\begin{align}
 \rho_\text{gcg}^{(\TT)}(x)&=\frac{\varXi^{(\LL)}(x)\,\varXi^{(\RR)}(x)}{\Lambda \Xi_\text{gcg}}\,e^{-\beta \Vext^{(\TT)}(x)}\,,\label{eq_rhoIgcT} \\
 \rho_\text{gcg}^{(\alpha)}(x)&=\int\mathrm{d}x_0\,\rho_\text{gcg}^{(\TT)}(x_0)\,\varrho^{(\alpha)}(x|x_0)\,.\label{eq_rhoIgcGEN}
\end{align}
Intriguingly, if the same external potential acts on all particles, the total density in an ordinary grand-canonical fluid (with partition function $\Xi$) is equal to the 
ordered density $\rho_\text{gcg}^{(\TT)}$ of the tagged particle (up to the different normalization).
This becomes evident from comparing the representation of the former given in Ref.~\cite{percus2} to Eq.~\eqref{eq_rhoIgcT}.
Finally, the ggg ensemble constitutes the basis for a full DFT treatment of the ordering problem. 
To achieve this, it is necessary to properly account for the order-preserving potential, Eq.~\eqref{eq_WNalpha} in a functional on the many-body level.
In such an ensemble it is, however, not possible to define conditional densities in the spirit of the above considerations.

\begin{figure}[t]
\includegraphics[width=0.45\textwidth] {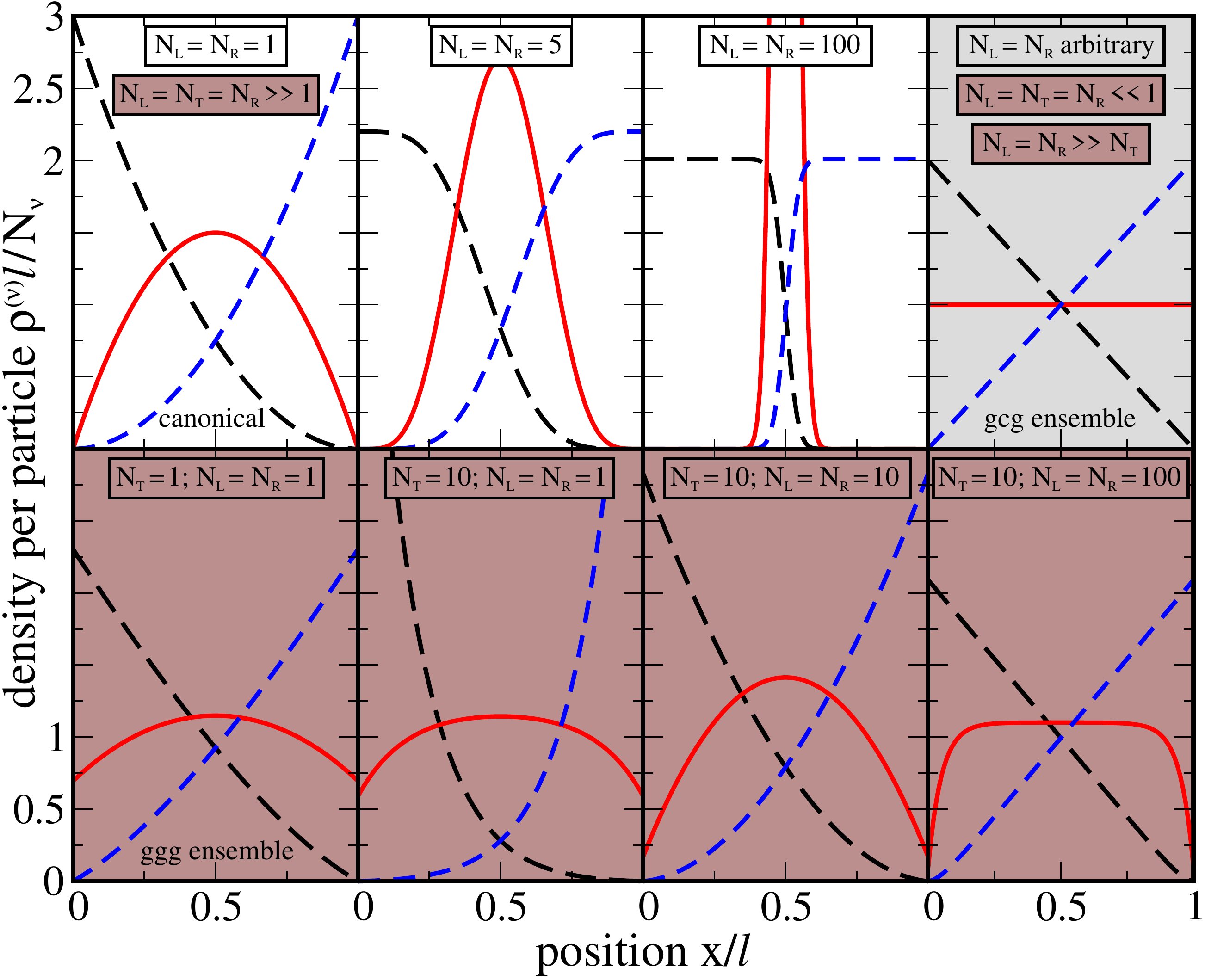} 
\caption{ Comparison of normalized equilibrium density profiles for ordered point particles in the different ensembles defined in Secs.~\ref{sec_c} and~\ref{sec_gc}.
Shown are the density profiles $l\rho^{(\nu)}(x)/N_\nu$ of the three components $\nu$ 
in a slit of {arbitrary unit} length $l$ normalized by the particle numbers $N_\nu$ (understood as an average value if $\nu$ is treated grand-canonically).
The tagged species ($\nu\!=\!{\TT}$) and the species to its left ($\nu\!=\!{\LL}$) and right ($\nu\!=\!{\RR}$) are shown as solid red, long-dashed black and dashed blue lines, respectively.
\textbf{Top:} canonical (ccc) ensemble (white background) for intrinsically fixed  $N_{\TT}\!=\!1$ and different $N_{\LL}\!=\! N_{\RR}$ as labeled
and gcg ensemble (on the right, gray background) for intrinsically fixed $N_{\TT}\!=\!1$ and arbitrary $N_{\LL}\!=\! N_{\RR}\!=\!(N-1)/2$, which all collapse on the same curves.
The additional labels with the brown background indicate the limits in the ggg ensemble, for which the same density profiles are recovered, cf.\ appendix~\ref{app_ideal}.
\textbf{Bottom:} ggg ensemble (brown background) for different $N_\nu$ as labeled.
\label{fig_EQUILIBRIUM}
}
\end{figure}

\subsection{Ordered distributions of point particles in different ensembles \label{sec_ensembles}}

Comparing the analytical density profiles of point particles in a slit of length $l$, calculated in appendix \ref{app_ideal}, we
illustrate in Fig.~\ref{fig_EQUILIBRIUM} the relation between the different ensembles 
and argue in how far there exists a ''thermodynamic limit``, in which the ensembles would become equivalent.
For this reason, we restrict ourselves to symmetric systems with an equal number $N_\alpha\equiv N_{\LL}\equiv N_{\RR}$ of host particles at each side of the tagged particle. 
As we here consider point particles, the density of each species $\nu$ is nonzero at any point in the slit.
{Note that without imposing the particle order at this stage, such a setup would represent three independent ideal-gas systems in a slit,
such that all curves would simply reduce to horizontal lines with $l\rho^{(\nu)}(x)/N_\nu\equiv1$.}

In the canonical case, cf.~Eq.~\eqref{eq_rhoIcanGEN}, the distribution of the tagged particle is clearly peaked in the middle of the slit.
The peak becomes sharper for increasing $N$ as the distribution of the host particles develops a pronounced plateau.
 For an infinite system, the variance of the tagged-particle profile scales like $l^2/N$.
In stark contrast, the tagged-particle profile in the gcg ensemble, cf.~Eq.~\eqref{eq_rhoGCG} or Eqs.~\eqref{eq_rhoIgcT} and~\eqref{eq_rhoIgcGEN}, is both constant and independent of $N$, such that the variance scales like $l^2/12$.
The distribution of the other species is always a linear function with $N$ as a multiplicative factor.
This means that there is no limit in which this ensemble with fluctuating particle numbers reduces to the canonical one.

Turning now to the ggg ensemble, cf.~Eq.~\eqref{eq_rhoGGG}, and generalizing the notion of a tagged particle to a species holding on average $N_{\TT}$ particles,
several scenarios can occur.
Assuming first an equal value of the number $N_\nu\equiv N_\alpha\equiv N_{\TT}$ of particles in each species (which does not mean that the chemical potentials $\mu_\nu$ are equal),
we find that the densities per particle in the many-particle case $N_\nu\gg1$ are the same as in  the ccc ensemble with $N_\alpha=1$.
In turn, the low-density case $N_\nu\ll1$ resembles the gcg situation.
The gcg densities per particle are more generally recovered whenever $N_\alpha\gg N_{\TT}$,
which implies that in the limit $N_\alpha\rightarrow\infty$ {(while $N_{\TT}=1$ is fixed)} the gcg and ggg ensembles are equivalent.
Finally, in the opposite case $N_\alpha\ll N_{\TT}$, the density of the tagged species becomes again constant,
while the other particles are trapped close to the boundary.
The (rather unappealing) limit $N_{\TT}\rightarrow\infty$ and $N_\alpha=1$ limit is now equivalent to the ordered canonical ensemble
(and would also be equivalent to a cgc ensemble not considered here).
These trends are apparent from the bottom plots in Fig.~\ref{fig_EQUILIBRIUM} and are explicitly evaluated in appendix~\ref{app_ideal}.

\section{Adiabatic dynamics of ordered particles \label{sec_OPD}}

Having understood the implications of an ordered equilibrium in statistical mechanics,
we exploit this framework in a more general context
to explore the tagged-particle dynamics in an overdamped single-file system.
To be able to do so, we employ the infamous adiabatic approximation,
assuming that the density profiles at any time can be written in the form of Eq.~\eqref{eq_rhoIcanGEN}, i.e., as if the system was in equilibrium.

\subsection{From particle-conserving dynamics (PCD) to order-preserving dynamics (OPD)}

The fundamental idea behind PCD \cite{delasherasbrader2016,schindlerproject} is to replace the grand-canonical expression of the underlying intrinsic Helmholtz free energy functional $\mathcal{F}[\{\rho^{(\nu)}(x)\}]$ in the standard (adiabatic) DDFT equation \cite{archer_evans,marconi_tarazona}
\begin{align}
 \frac{\partial \rho^{(\nu)}(x,t)}{\partial t}= \beta D_0\frac{\partial}{\partial x}\,\rho^{(\nu)}\,\frac{\partial}{\partial x}\left({\frac{\delta\mathcal{F}}{\delta\rho^{(\nu)}}}+\Vext^{(\nu)}(x)\right)
 \label{eq_DDFTgc}
\end{align}
for the time-dependent one-body density $\rho^{(\nu)}(x,t)$ with a canonical one, where $D_0$ is the short-time Brownian diffusion coefficient.
While the particle numbers $N_\nu$ of the different components, i.e., the integral of the densities, {always remain the same for all times} $t$,
the key difference {between DDFT and PCD} is the interpretation and explicit form of $\rho^{(\nu)}(x,t)$.
For a grand-canonical free energy, {this quantity} has to be understood as an average over the canonical densities $\rho_N^{(\nu)}(x,t)$ of different systems with {fixed integer values of $N_\nu$}.

The required canonical functional
\begin{align}
 \!\!\! \beta \mathcal{F}_{N}\big[\{\rho_N^{(\nu)}\}\big]=-\ln \CPF_{N}-\sum_{\nu}\int\upd x\,\rho_N^{(\nu)}(x)\beta V^{(\nu)}(x) \!\!
 \label{eq_Fcan}
\end{align}
can be formally defined by subtracting the extrinsic contribution from the total free energy $-\ln \CPF_{N}$ of an equilibrium system in the given external potentials $\Vext^{(\nu)}(x)$.
Here, $\rho_N^{(\nu)}(x)$ are understood as the one-body densities, which are in canonical equilibrium in an imaginary system with the \textit{generating} external potentials $V^{(\nu)}(x)$ replacing the actual ones $\Vext^{(\nu)}(x)$.
Eq.~\eqref{eq_Fcan} is thus to be interpreted as the defining equation of the generating potentials,
since we formally search the value of $\mathcal{F}_{N}$ for the given canonical \textit{target} densities $\rho_N^{(\nu)}(x)$.
If and only if all $V^{(\nu)}(x)=\Vext^{(\nu)}(x)$, then the true equilibrium densities $\rho_N^{(\nu)}(x)=\rho_{N,\text{eq}}^{(\nu)}(x)$ are considered for which $\mathcal{F}_{N}$ becomes minimal.
Note that $\CPF_N$ is thus a functional of $\rho_{N,\text{eq}}^{(\nu)}(x)$ only and not of the general target densities.
Substituting now Eq.~\eqref{eq_Fcan} into Eq.~\eqref{eq_DDFTgc} yields the particle-conserving evolution equation
\begin{align}
 \frac{\partial \rho_N^{(\nu)}(x,t)}{\partial t}= \beta D_0\frac{\partial}{\partial x}\,\rho_N^{(\nu)}\,\frac{\partial}{\partial x}\left(\Vext^{(\nu)}(x)-V_\text{ad}^{(\nu)}(x,t)\right)
 \label{eq_DDFTgen}
\end{align}
for the canonical densities $\rho_N^{(\nu)}(x,t)$,
where $V_\text{ad}^{(\nu)}(x,t)$ is identified with the generating potential of $\rho_N^{(\nu)}(x,t)$ at a given time $t$,
such that $\lim_{t\rightarrow\infty}V_\text{ad}^{(\nu)}(x,t)=\Vext(x)$.
What is now left to be specified is how we conceive the notion of an equilibrium system, which defines $V^{(\nu)}(x)$ and $V_\text{ad}^{(\nu)}(x,t)$.

The modified DDFT equation, Eq.~\eqref{eq_DDFTgen}, is generically valid for any rule by which the generating external potentials are constructed.
To see this, we consider the iteration scheme \cite{schindlerproject,delasherasbrader2016}
\begin{align}
 \beta V^{(\nu)}_{n}(x)=\beta V^{(\nu)}_{n-1}(x)-\ln{\rho_N^{(\nu)}}(x)+\ln\rho^{(\nu)}_{n-1}(x)\,,
 \label{eq_Vit}
\end{align}
where the potentials $V^{(\nu)}_{n}(x)$ are updated in each step $n$.
After a sufficient number of steps, $V^{(\nu)}_{n}(x)$ have converged to the potentials $V^{(\nu)}(x)$ generating the given $\rho_N^{(\nu)}(x)$.
Strictly speaking, it is not necessary to associate these target densities with any statistical ensemble.
The crucial point that specifies the nature and explicit form of $V^{(\nu)}(x)$ is rather how, i.e., 
with respect to which ensemble, in iteration step $n$, the densities $\rho^{(\nu)}_{n-1}(x)$ are obtained from $V^{(\nu)}_{n-1}(x)$ of the previous step.
For example, if $\rho^{(\nu)}_{n-1}(x)$ was determined from the minimization of a grand-canonical DFT, then
Eq.~\eqref{eq_DDFTgen} would be based on a grand-canonical rule and thus be equivalent to the DDFT from Eq.~\eqref{eq_DDFTgc}~\cite{noteGC}.

Turning now to a canonical system, we have discussed in Sec.~\ref{sec_statmech}
that we can distinguish between ordinary and ordered canonical equilibrium densities, 
for which the canonical partition function $\CPF_N$ is identical.
This means that {Eq.~\eqref{eq_Vit} will converge to a different generating potential, if particle order is enforced by using the ordered densities given by Eq.~\eqref{eq_rhoIcanGEN} 
or, more generally, by inverting the conditional DFT described in appendix~\ref{app_DDFT}.}
{As a result}, also the {functional dependence of the free energy on the density, implicit in Eq.~\eqref{eq_Fcan}, is different in an ordered system.
Therefore, the corresponding} PCD based on Eq.~\eqref{eq_DDFTgen} evolves into OPD,
 a theory, which preserves the particle order between the different species in one dimension.
As already pointed out in Ref.~\cite{schindlerproject}, the main distinction between these two approaches is their different equilibrium limit.
In the remainder of this work, we explore the further differences in the dynamical behavior.

\subsection{PCD vs.\ OPD for hard rods \label{sec_comparison}}

In the following, we discuss the dynamics of $N=2$ and $N=3$ hard rods of length $\sigma$ confined to a slit of length $l=4.9\sigma$, 
for which the ordered densities from Eq.~\eqref{eq_rhoIcanGEN} can be efficiently calculated.
 Note that for $N=2$ it does not matter, whether the left or right particle is interpreted as the tagged particle.
Moreover, setting in general, e.g., $N_{\LL}=N$ and $N_\TT=N_\RR=0$,
all information about order is sacrificed and OPD corresponds to PCD for a single species \cite{delasherasbrader2016}.
An intriguing result from Ref.~\cite{schindlerproject} {is that this collective version of PCD
is not generally recovered upon 
constructing a total density $\sum_\nu\rho_N^{(\nu)}(x,t)$
by adding up the individual profiles 
found in the particle-resolved version of PCD}.

Having introduced OPD which overcomes the fundamental drawback of missing particle order in PCD, 
we now seek an answer to the three remaining questions.
In how far does OPD constitute an improvement over the PCD results for single particles from Ref.~\cite{schindlerproject}?
Is the total density of both theories equivalent?
How are the predictions of OPD affected if one species holds more than one particle?

To {answer these questions, we also} compare our results to Brownian dynamics (BD) simulations.
We choose similar setups as in Ref.~\cite{schindlerproject},
where the particles are initially equilibrated in harmonic potentials $V_\text{ad}^{(\nu)}(x,0)=k^{(\nu)}(x-x_\text{h}^{(\nu)})^2/2$
centered at $x_\text{h}^{(\nu)}$ with the constants $k^{(\nu)}$
and then released at $t=0$.
{At all times, the particles are confined between two hard walls at $x=0$ and $x=l$.}
Recall that on the single-particle level, the initial profiles of PCD, do not exactly match those from BD and OPD, which both accurately represent the ordered equilibrium in given $V_\text{ad}^{(\nu)}(x,0)$.
Time is measured in terms of the Brownian time $\tau_\text{B}=\sigma^2/D_0$.

\begin{figure}[t]
\includegraphics[width=0.4425\textwidth] {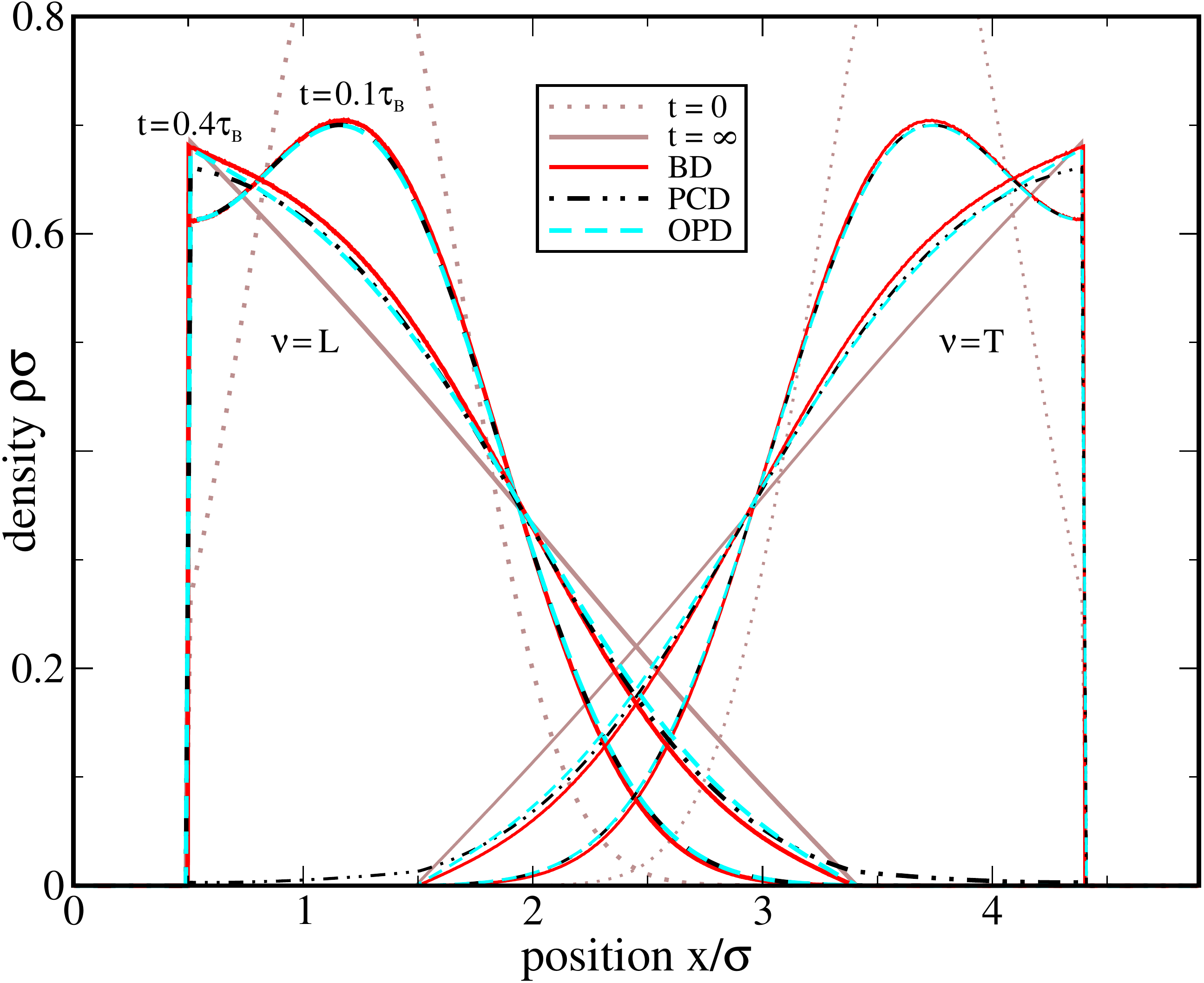} 
\caption{Individual density profiles $\rho_N^{(\nu)}(x,t)$ of $N\!=\!2$ hard rods of length $\sigma$ on a line of length $l\!=\!4.9\sigma$ for distinguished initial trapping 
with $k^{(\LL)}\!=\!k^{(\TT)}\!=\!5/(\beta\sigma^2)$, $x_\text{h}^{(\TT)}\!=\!l/4$ and $x_\text{h}^{(\TT)}\!=\!3l/4$ (species $\nu\!=\!{\LL}$ and $\nu\!=\!{\TT}$ as labeled).
We compare BD and PCD results as in Ref.~\cite{schindlerproject} to OPD {at times $t=0.1\tau_\text{B}$ and $t=0.4\tau_\text{B}$ (as labeled).
The common initial ($t=0$) and final ($t=\infty$) profiles of OPD and BCD are drawn as a reference.
A more detailed figure for this setup, including different time steps, is attached to this manuscript as Fig.~\ref{fig_T2figCL}.}
\label{fig_T2fig}
}
\end{figure}

\subsubsection{Individual profiles for distinguished initial trapping \label{sec_compareINDIVIDUAL}}

For two particles starting at distinct locations, as shown in Fig.~\ref{fig_T2fig}, the initial PCD profiles of each particle are highly accurate \cite{schindlerproject}.
Hence, at short times, there is no significant difference between the considered approaches.
Both the decrease of the maximal density at the central peaks and the increase of the contact density at the wall occur slightly faster for OPD than for PCD. 
Compared to the exact BD, the decay of the peaks is too fast in both theories, a known consequence of the adiabatic assumption \cite{marconi_tarazona,penna2006}.
In turn, there is a perfect match between OPD and BD, regarding the accumulation at the system boundary,
while PCD falls slightly behind for larger times.
The superiority of OPD compared to PCD becomes most apparent when equilibrium is approached, as anticipated from the ordered ensemble underlying OPD.
In particular, the points beyond which the density profiles in OPD constantly vanish fully agree with the BD result, while this happens in PCD only at the system boundaries due to unphysical particle mixing.

\begin{figure}[t]
\includegraphics[width=0.4425\textwidth] {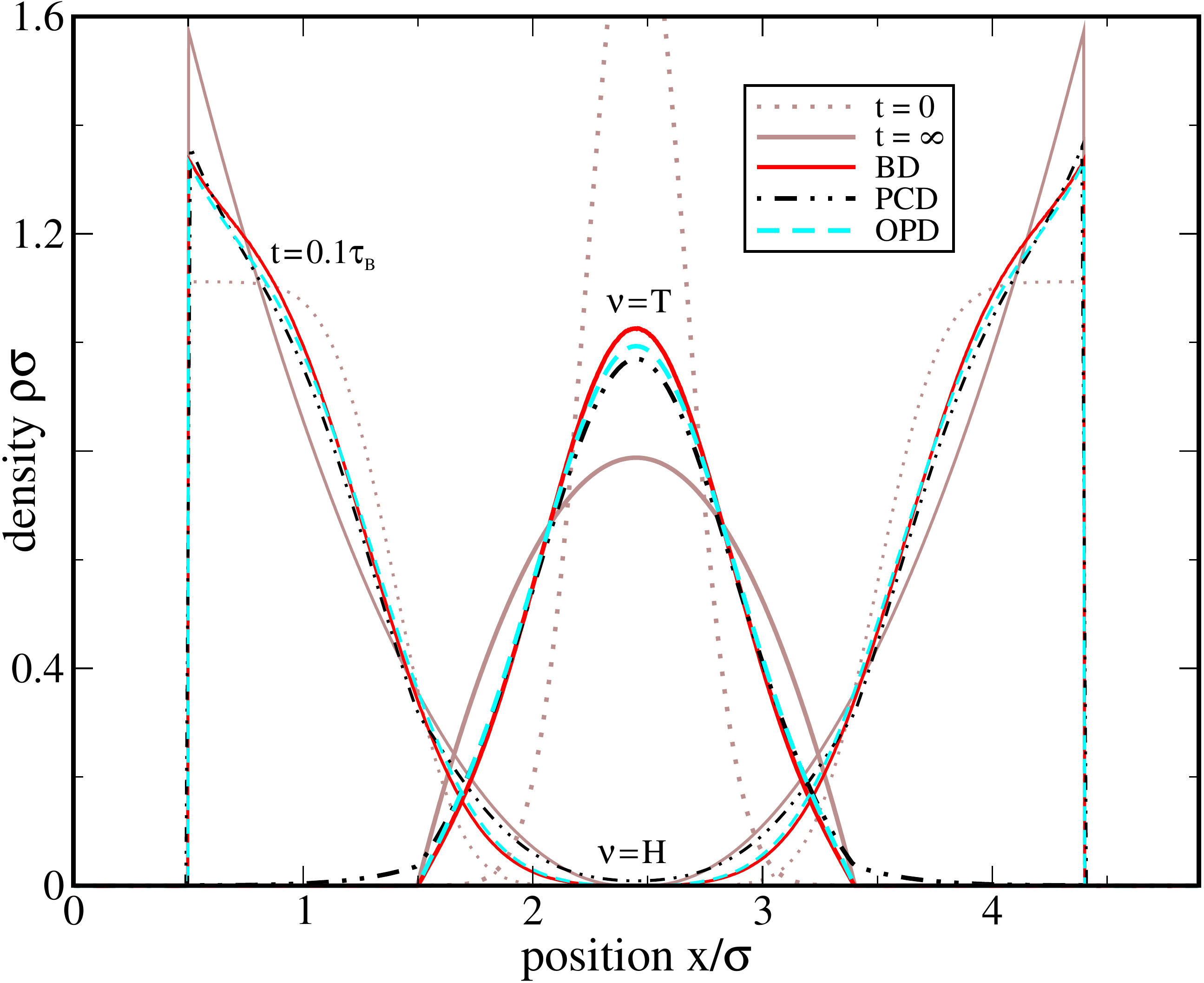} 
\caption{Individual density profiles as in Fig.~\ref{fig_T2fig} but for $N\!=\!3$ hard rods with $k^{(\TT)}\!=\!20/(\beta\sigma^2)$, $x_\text{h}^{(\TT)}\!=\!l/2$ and $k^{(\LL)}\!=\!k^{(\RR)}\!=\!0$ (species $\nu\!=\!{\LL}$ and $\nu\!=\!{\RR}$ are joined to a single species $\nu\!=\!{\HH}$).
{Shown are the data at time $t=0.1\tau_\text{B}$.
A more detailed figure for this setup, including different time steps, is attached to this manuscript as Fig.~\ref{fig_T3figCL}.}
\label{fig_T3fig}
}
\end{figure}

A similar situation occurs for three particles, as shown in Fig.~\ref{fig_T3fig}.
The peaks of the OPD profiles decay faster than for the reference BD results, but the behavior at the boundary is again the same.
In this case, the disordered nature of PCD is already evident upon initialization through the deviations from the exact BD and OPD profiles at $t=0$ \cite{schindlerproject} and,
since the system is more densely packed, the mixing of the profiles over time happens much faster.

\subsubsection{Total profile for common initial trapping \label{sec_compareTOTAL}}

As a second step, we compare OPD and PCD on the level of the total density of all particles.
We thus switch to an unbiased setting with the initial trapping potential $V_\text{ad}(x,0)\equiv V_\text{ad}^{(\nu)}(x,0)$ acting on all particles of any species $\nu$ in the same way.
This setup allows us to test both theories against BD for more extreme initial conditions with stronger correlations and a lower probability to find any particle close to the wall.
{Note that} the species-resolved PCD consistently yields he same result $\rho_N^{(\nu)}\propto\rho_N$ as PCD for a single species at all times.
Therefore, both the initial profile $\rho_N(x,0)$ and the long-time equilibrium limit $\rho_N(x,\infty)$ of the total density are exact in PCD (as they are always in OPD).

\begin{figure}[t]
\includegraphics[width=0.4425\textwidth] {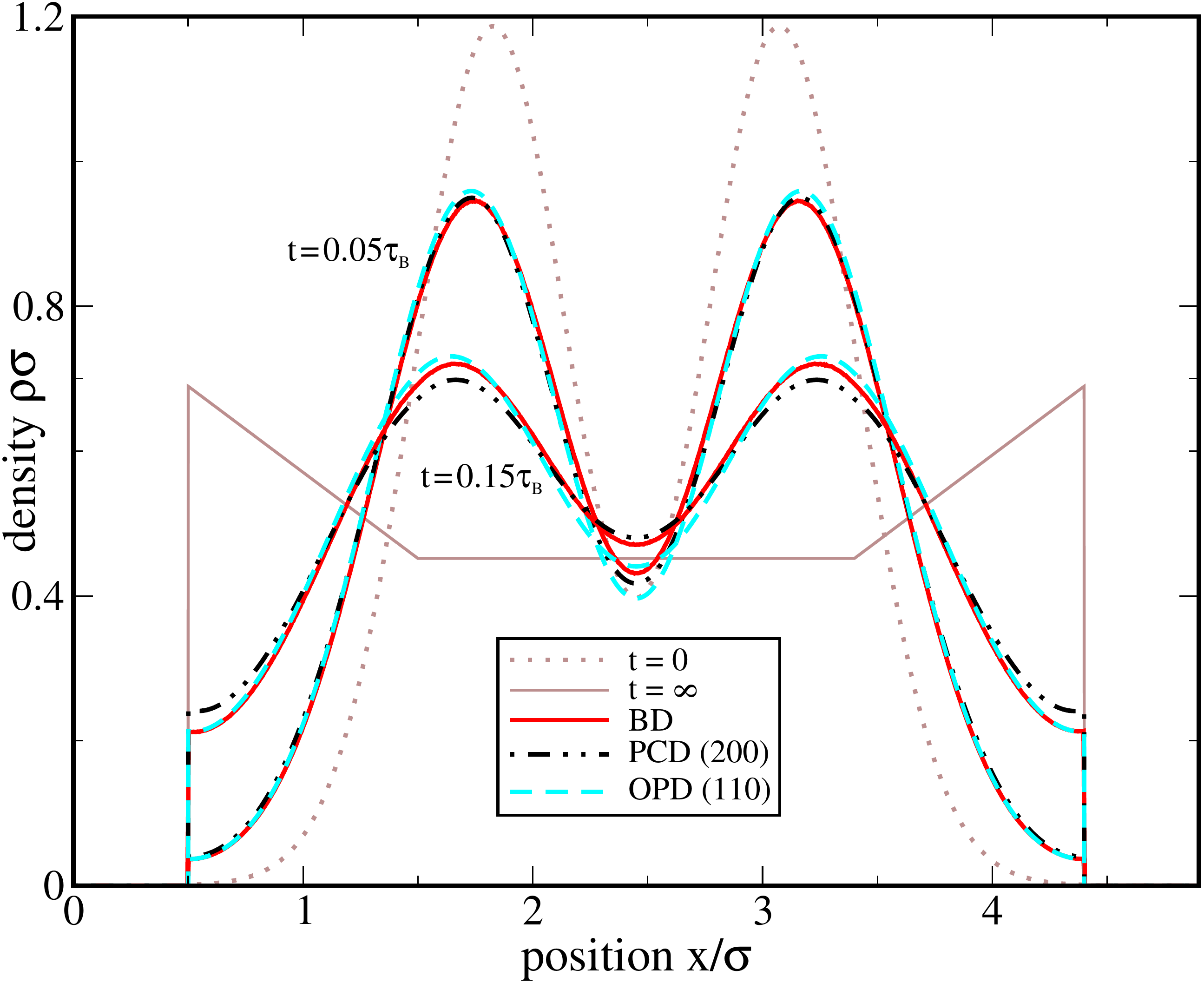} 
\caption{
Total density profiles $\rho_N(x,t)$ of $N\!=\!2$ hard rods of length $\sigma$ on a line of length $l\!=\!4.9\sigma$ for common initial trapping 
with $k^{(\nu)}\!=\!5/(\beta\sigma^2)$ and $x_\text{h}^{(\nu)}\!=\!l/2$ using different approaches according to the legend.
The numbers $(N_{\LL}N_{\TT}N_{\RR})$ in brackets indicate the numbers $N_\nu$ of particles in each component $\nu$ of OPD, where PCD is formally equal to OPD if all particles belong to the same species.
{The common initial ($t=0$) and final ($t=\infty$) profiles of all approaches are drawn as a reference
and the intermediate steps are at $t=0.05\tau_\text{B}$ and $t=0.15\tau_\text{B}$ (as labeled). 
A more detailed figure for this setup, including different time steps, is attached to this manuscript as Fig.~\ref{fig_T2midCL}.}
\label{fig_T2mid}
}\end{figure}

\begin{figure}[t]
\includegraphics[width=0.4425\textwidth] {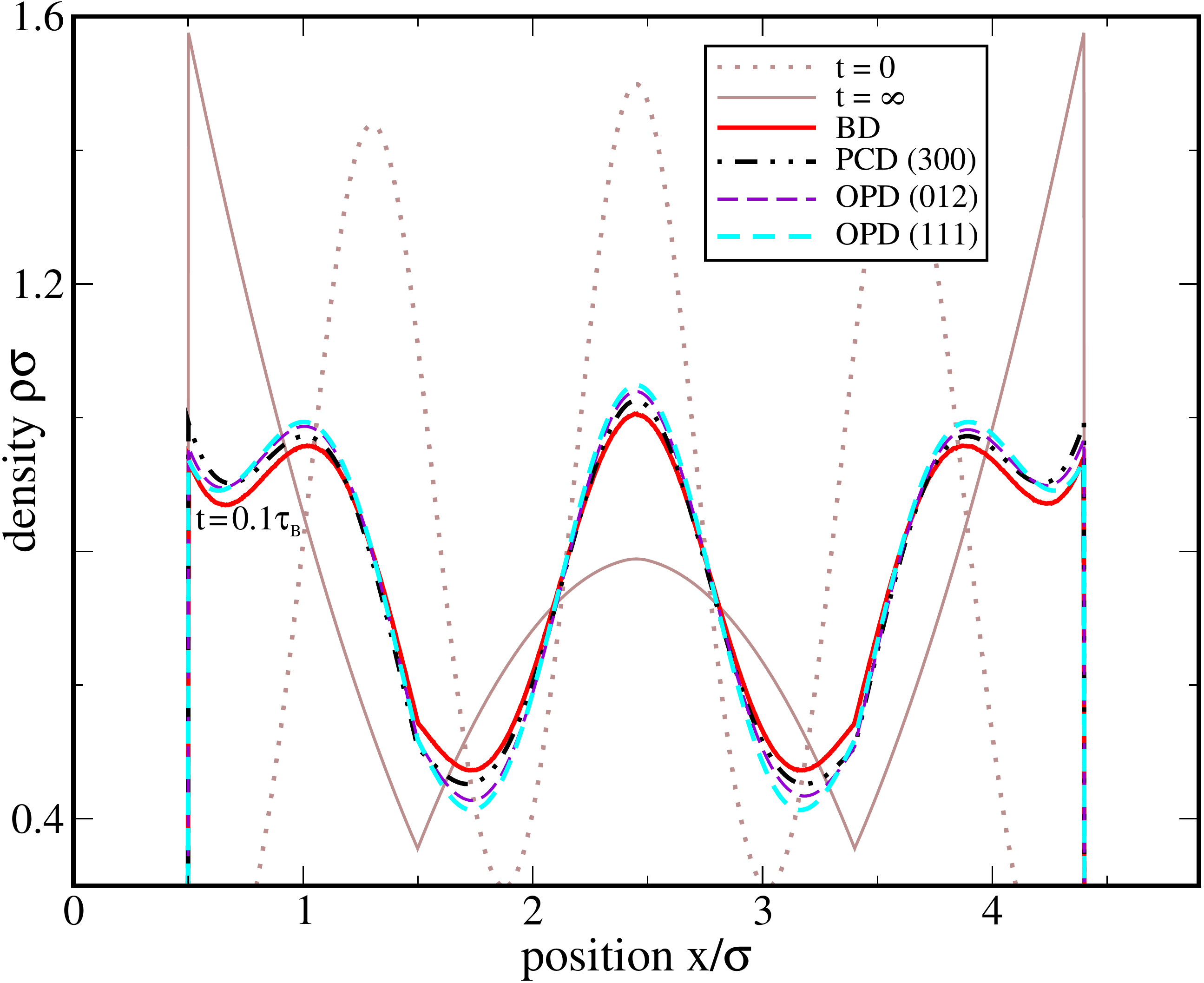} 
\caption{Total density profiles as in Fig.~\ref{fig_T2mid} but for $N\!=\!3$ hard rods. One additional set of OPD data is included considering the two particles on the right as members of a single species. 
 Joining the left and right particle to a new host species with $\rho_N^{(\HH)}=\rho_N^{(\LL)}+\rho_N^{(\RR)}$ yields the same result as for full OPD with three species.
 {Shown are the data at time $t=0.1\tau_\text{B}$.
A more detailed figure for this setup, including different time steps, is attached to this manuscript as Fig.~\ref{fig_T3midCL}.}
\label{fig_T3mid}
}\end{figure}

From the symmetric setups in Fig.~\ref{fig_T2mid} and Fig.~\ref{fig_T3mid},
we deduce the general behavior that OPD predicts a slower decay of the density peaks compared to BD.
{This observation stands} in contrast to the case depicted in Fig.~\ref{fig_T2fig} and Fig.~\ref{fig_T3fig},
{where a larger separation of the particles in the initial state constitutes the main difference.
PCD predicts a similar behavior for $N=3$ in Fig.~\ref{fig_T3mid}, while for $N=2$ in Fig.~\ref{fig_T2mid} this is only the case in the early dynamics and PCD overtakes BD after a short time}.
This means that not only the time evolutions of the total density predicted by OPD and PCD differ, but they also do so in a qualitatively
 different way when compared with the reference BD scenario.
 The most striking observation discussed in Sec.~\ref{sec_compareINDIVIDUAL} turns out to be, in fact, a general result: 
  OPD predicts the correct contact value of the density at the system boundary, while there are strong deviations in the case of PCD.

Considering further in Fig.~\ref{fig_T3mid} a partial version of OPD for $N=3$ particles {but} with only two {different} species, we observe a dynamical behavior interpolating between PCD and three-species OPD.
As the first species holds $N_{\TT}=1$ particles and the second one $N_{\RR}=2$,
the total density is not symmetric with respect to $x=l/2$
and its values on the left-hand side are closer to those of full OPD than on the right-hand side.
Treating two {(or more)} particles as members of the same species thus amounts, in general, to an {additional} approximation.
Tagging, however, the central particle and formally joining the other two host particles to one component by calculating $\rho_N^{(\HH)}=\rho_N^{(\LL)}+\rho_N^{(\RR)}$
does not make a difference in Fig.~\ref{fig_T3mid}, since $\rho_N^{(\HH)}(l/2,t)\equiv0$ for the system length considered here.
Increasing $l$ by $\sigma$, this is no longer the case, but still does not result in any noticeable deviation from the full OPD data.
{This point and further conclusions for such a larger system are detailed in appendix~\ref{sec_compareTOTALasym}.}

\subsection{Single-file diffusion (SFD) of point particles}

We have seen that, leaving the effects of the adiabatic approximation aside, OPD provides an excellent account of the dynamics in small confined systems, 
but also that it makes a difference if some particles are grouped together within the same species. 
 Now we push this three-component approach for OPD to more complex nonequilibrium scenarios by gradually taking the limits of an infinite system and an infinite number of host particles.
In particular, we wish to clarify whether we can predict some universal long-time behavior,
carrying the footprints of single-file diffusion (SFD), when tagging the central particle in a larger system.
Recall that, in a confined system, the exact equilibrium limit is recovered by construction, 
while for a more general diffusion problem in an infinite system equilibrium is practically never reached.

As we are particularly interested in the time evolution of the mean-square displacement (MSD) $\Delta x^2(t)$,
the size of the particles is only of minor importance.
We thus restrict ourselves in this study to point particles, whose tagged-particle distributions are exactly known for a conserved number of particles \cite{roedenbeckhahnSFD1998,kumar2008}.
Here, we also consider an OPD scenario based on the gcg ensemble, where the (equal) numbers $N_\alpha=\int\upd x\,\rho_\text{gcg}^{(\alpha)}(x,t)$ of host particles are fixed on a grand-canonical-type average.
There {are two} main reasons to do so.
First, we can explore the relation between the different ensembles in a more general (dynamical) setup than in Sec.~\ref{sec_ensembles}.
Second, this rather artificial ensemble seems more natural from a DDFT point of view,
since more general calculations of this type for a large number of interacting particles usually require grand-canonical DDFT methods.
We distinguish between the two scenarios considered by explicitly speaking of an OPD with fluctuating particle numbers {(or fluctuating OPD)} if we do not mean the (canonical) OPD exclusively used hitherto.

 Similar to previous implementations of the dynamical test-particle approach, 
we initialize the system in a state, identified with the time $t=10^{-5}\tau_\text{B}$ by choosing a harmonic trap with $k^{(\TT)}\!=\!5\cdot10^{4}/(\beta\sigma^2)$ and $x_\text{h}^{(\TT)}\!=\!0$.
At this early stage, the exact density of the tagged particle is still nearly Gaussian and follows the ideal diffusion law for the chosen concentrations \cite{roedenbeckhahnSFD1998}.
To be able to swipe through many different time regimes, we gradually decrease the spatial and temporal resolution, as all functions become smoother with increasing time.
We always assume {an initial state where} the host particles are homogeneously distributed with the concentration $c=2N_\alpha/l$. 
Within any approach, the MSD 
\begin{align}
 \Delta x^2(t)=\int\upd x\, x^2\rho^{(\TT)}(x,t)
 \label{eq_MSD}
\end{align}
follows as the second moment of the tagged-particle density.

\begin{figure}[t]
\includegraphics[scale=0.3] {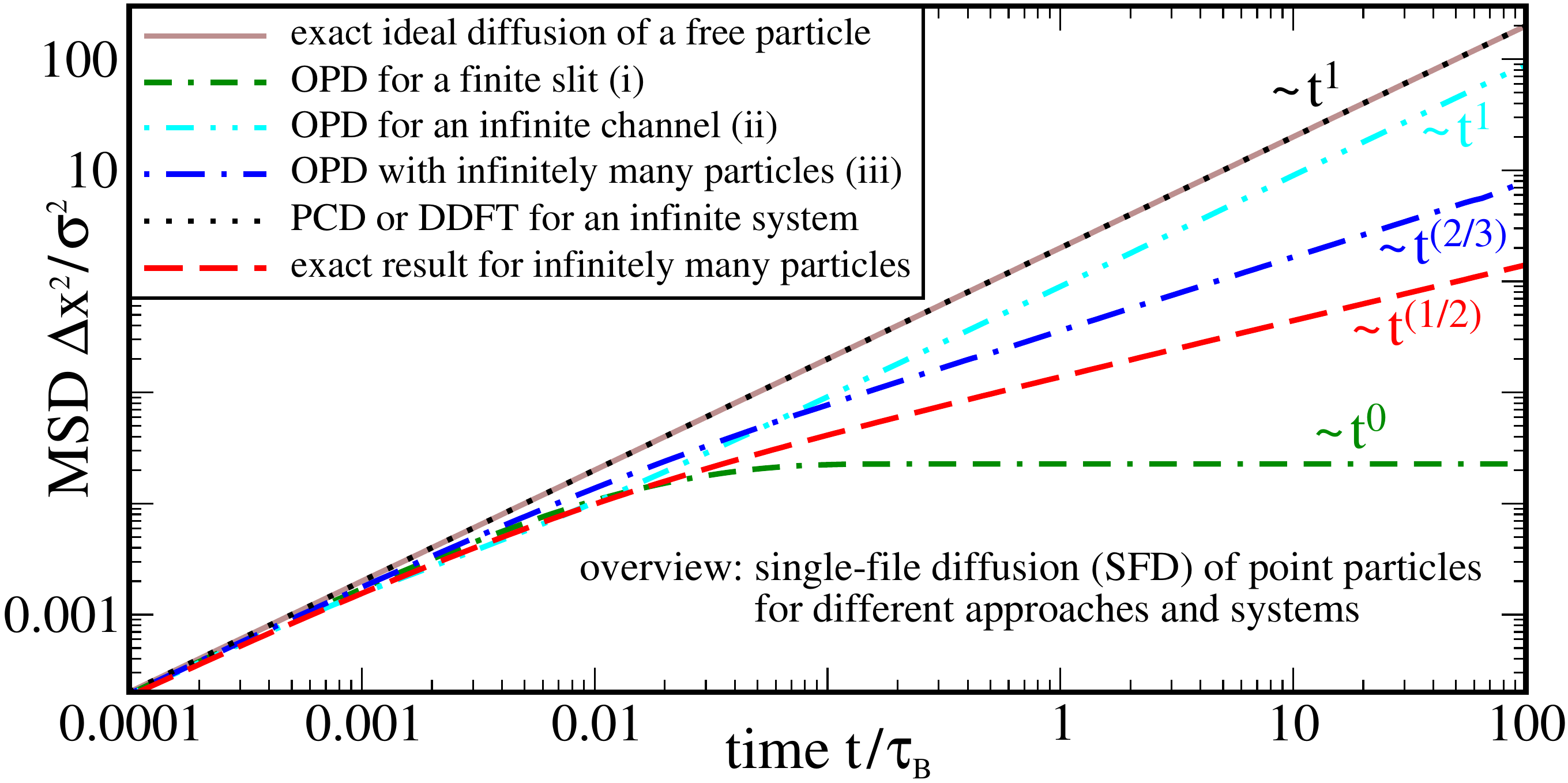} 
\caption{Summary of the results for the mean-square displacement (MSD) of a tagged particle, calculated according to Eq.~\eqref{eq_MSD}, in single-file diffusion (SFD) of point particles. 
Here $\sigma$ denotes the same arbitrary unit length as used in Brownian time $\tau_\text{B}\!=\!\sigma^2/D_0$.
OPD predicts (i) the correct maximal MSD in a finite slit, cf.\ Sec.~\ref{sec_SFDbox}, 
(ii) the correct reduced long-time self diffusion coefficient for a finite number of host particles in an infinite channel, cf.\ Sec.~\ref{sec_SFDfin},
and (iii) subdiffusive behavior for an infinite number of host particles with the exponent $2/3$, cf.\ Sec.~\ref{sec_SFDlim}.
For comparison, the exact subdiffusive behavior with exponent $1/2$, extracted from the density profiles calculated in Ref.~\cite{roedenbeckhahnSFD1998},
and the ideal diffusion law $\Delta x^2(t)\!=\!2D_0t$ in an open system are shown.
 The latter also corresponds to the generic prediction of PCD (or DDFT) for an ideal gas without particle order.
The concentration of host particles is $c\!=\!8/\sigma$ in all cases and $N_\alpha\!=\!4$ if not taken to infinity.
\label{fig_MSDsum}
}\end{figure}

The results for different approaches and setups are summarized in Fig.~\ref{fig_MSDsum} and compared in the following
for their dependence on the number $N_\alpha$ and (initial) {concentration} $c$ of the host particles,
as well as, the role of the adiabatic approximation and chosen ensemble.
Our key observations are that OPD correctly reproduces the long-time self-diffusion coefficient for a finite number of particles and
 gives rise to subdiffusive behavior in the limit of infinite particle number and system size.
 In the latter case, 
  the MSD from both OPD scenarios appears to converge to the same curve, which behaves like
 \begin{align}
  \Delta x^2(t)=\sqrt{2}\left(\frac{D_0t}{c}\right)^{2/3}
  \label{eq_MSDOPD}
 \end{align}
 for long times, i.e., we find a universal SFD exponent $2/3$.
  The deviation from the exact exponent $1/2$ is presumably a consequence of the involved approximations.
However, our results demonstrate that it is possible to predict deviations from the ideal diffusion law in the long-time limit using DDFT methods without empirical input.

\begin{figure}[t]
\includegraphics[scale=0.3] {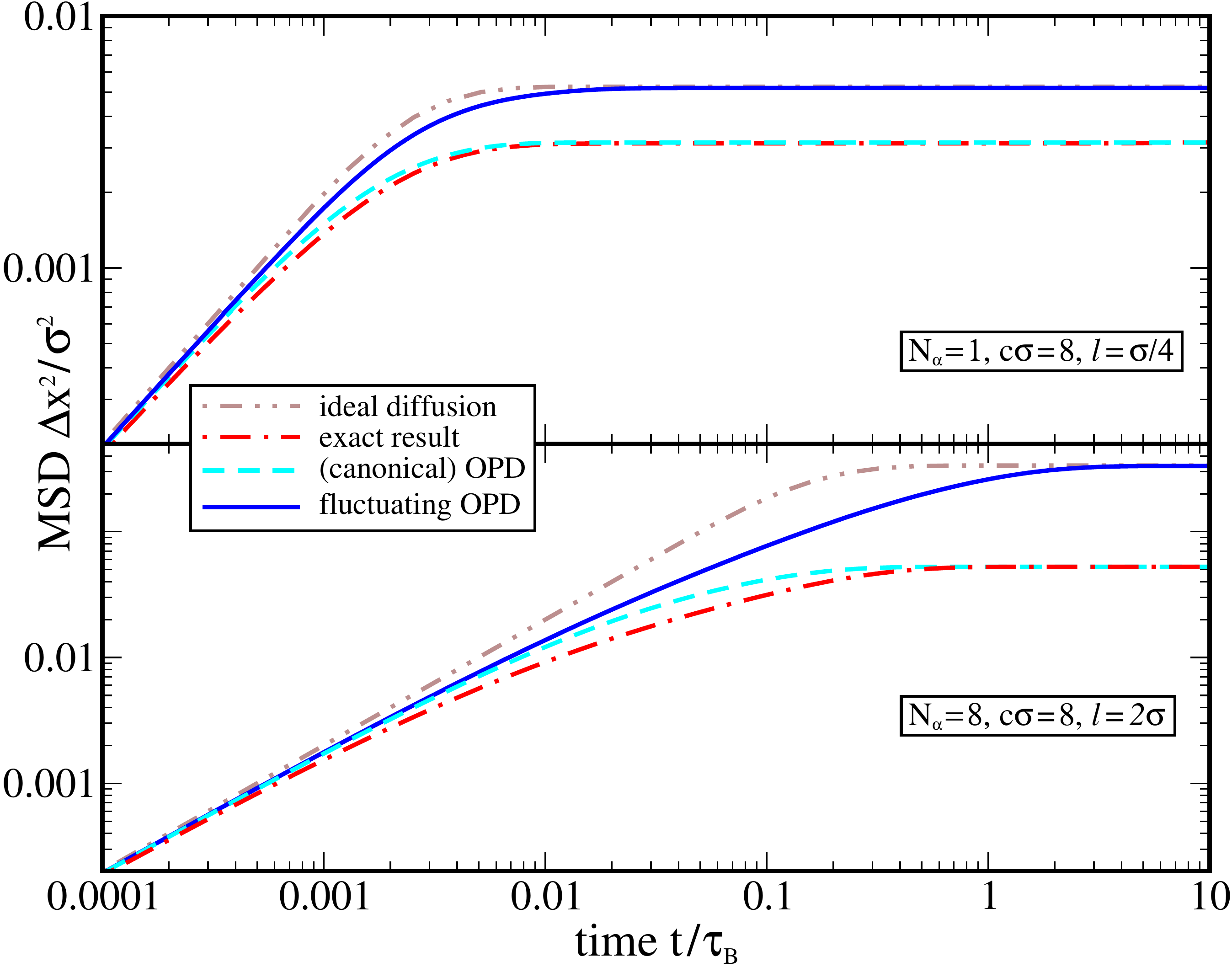} 
\caption{MSD of a tagged point particle in a confined system. 
{We} compare the OPD based on the ordered canonical ensemble {and} the fluctuating OPD based on the gcg ensemble 
{to the diffusion of a single (ideal) particle and exact results}. 
{According to the labels, we show two sets of data for box length $l$, number $N_\alpha$ of host particles and concentration $c\!=\!2N_\alpha/l$.
A more detailed figure for this setup, including further values of these parameters, is attached to this manuscript as Fig.~\ref{fig_MSDboxCL}.}
\label{fig_MSDbox}
}\end{figure}

\subsubsection{Finite system \label{sec_SFDbox}}

We start our detailed discussion by comparing the MSD in Fig.~\ref{fig_MSDbox} for a setup
similar to the one considered in Fig.~\ref{fig_T3fig}, but with a practically perfect initial location of the tagged particle.
The plateau value reached at large times is {always} determined by the equilibrium distribution for the given system size $l=2N_\alpha/c$.
For that reason, the OPD with fluctuating particle numbers, generally reaches a larger maximal MSD, cf.\ the different profiles in Fig.~\ref{fig_EQUILIBRIUM}.
 In fact, this limit is the same as for a single particle, which means that the fluctuating OPD does not improve much over PCD in this respect.
Compared to the ideal diffusion {(and PCD)}, the presence of host particles is, however, reflected {in fluctuating OPD} by the longer time it takes to reach the raised plateau.

For $N_\alpha=1$ host particle at each side, the MSD predicted by canonical OPD is barely distinguishable from the exact curve, as the equilibrium limit matches perfectly.
Only the transition from the ideal diffusive behavior to the plateau value of the MSD occurs slightly slower, 
 which is consistent with the observation, made in Fig.~\ref{fig_T3fig}, of a faster decrease of the central peak in OPD compared to BD.
Increasing $N_\alpha$, the deviations in the transient regime become more and more pronounced.

\begin{figure}[t]
\includegraphics[scale=0.3] {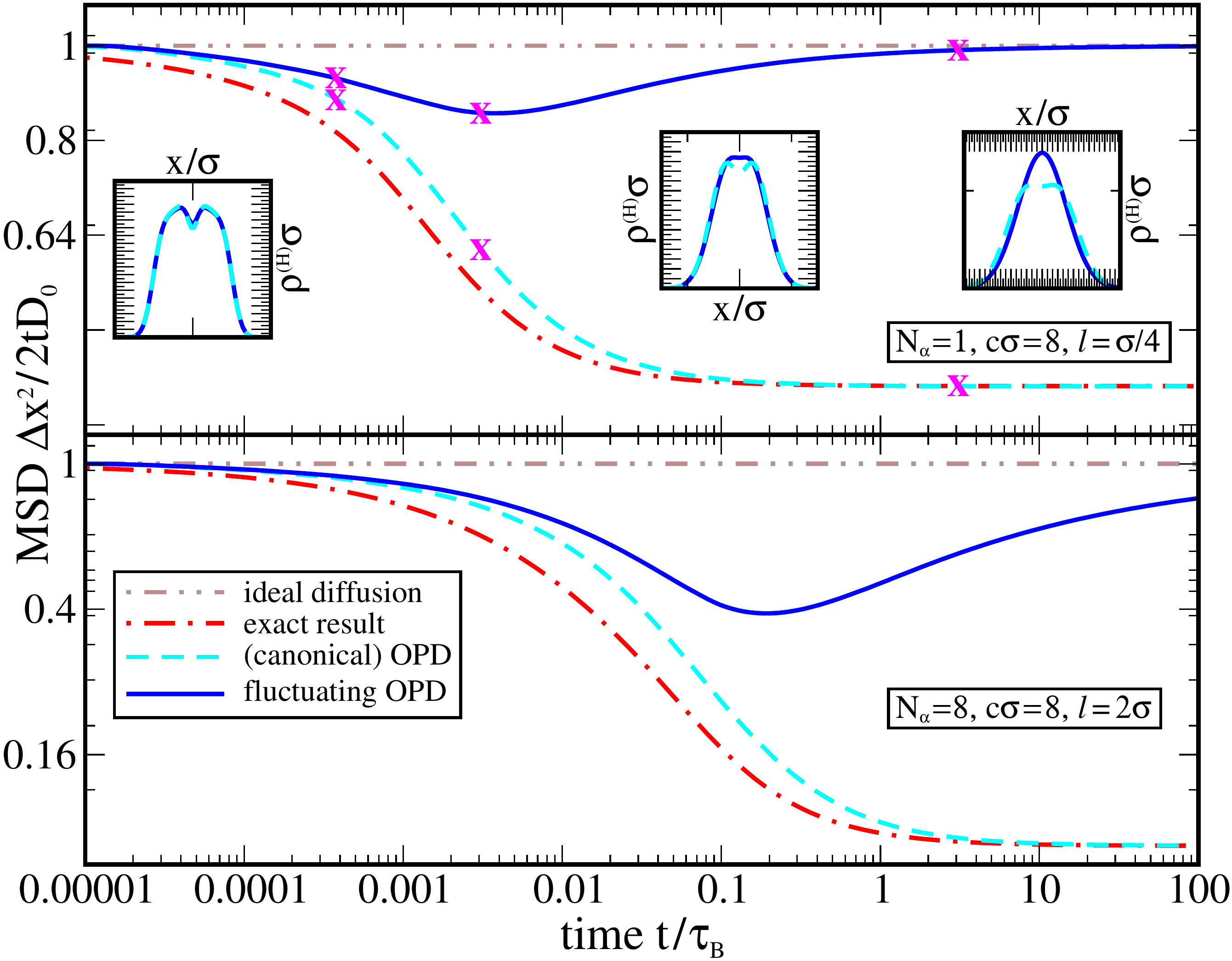} 
\caption{MSD of a tagged point particle as in Fig.~\ref{fig_MSDbox}, but for an open system with the host particles initially distributed homogeneously in a finite interval {of length $l\!=\!2N_\alpha/c$}.
Notice the different normalization of the vertical axis by the MSD $2D_0t$ of an ideally diffusing free particle ({brown} horizontal line at unity) to better resolve the differences at intermediate times.
The insets show the symmetric distribution $\rho^{(\HH)}(x,t)\!=\!\rho^{(\LL)}(x,t)+\rho^{(\RR)}(x,t)$ of host particles (representing the distinct part of the van Hove function \cite{hopkinsDTP2010}) 
with $\rho^{(\HH)}(\pm\infty,t)\!=\!0$ and axes as in Figs.~\ref{fig_T2fig} to~\ref{fig_T3large} with spacing 0.5 between major ticks.
 The data points for $t\!\approx\!0.00032\tau_\text{B}$, $t\!\approx\!0.0026\tau_\text{B}$ and $t\!\approx\!2.62\tau_\text{B}$ considered from left to right are marked by magenta {crosses}. 
 {A more detailed figure for this setup is attached to this manuscript as Fig.~\ref{fig_MSDfinCL}.}
  \label{fig_MSDfin}
}\end{figure}

\subsubsection{Infinite system with finite particle numbers \label{sec_SFDfin}}

Now we remove the system walls and let the host particles freely diffuse to the left and right, while using exactly the same homogeneous initial conditions within a finite region of length $2N_\alpha/c$. 
In this open system, the long-time behavior is again diffusive and we see in Fig.~\ref{fig_MSDfin} that OPD reproduces the exact dependence of the corresponding self-diffusion coefficient on the numbers of host particles.
{In particular, this diffusion coefficient does not depend on the initial particle} distribution, {for example, changing the concentration $c$ would only affect the crossover times between the different dynamical regimes \cite{SI}}.
 Again, the approach towards this limit takes longer in OPD, which becomes more and more apparent for a higher number of host particles.

 \begin{figure}[t]
\includegraphics[scale=0.3] {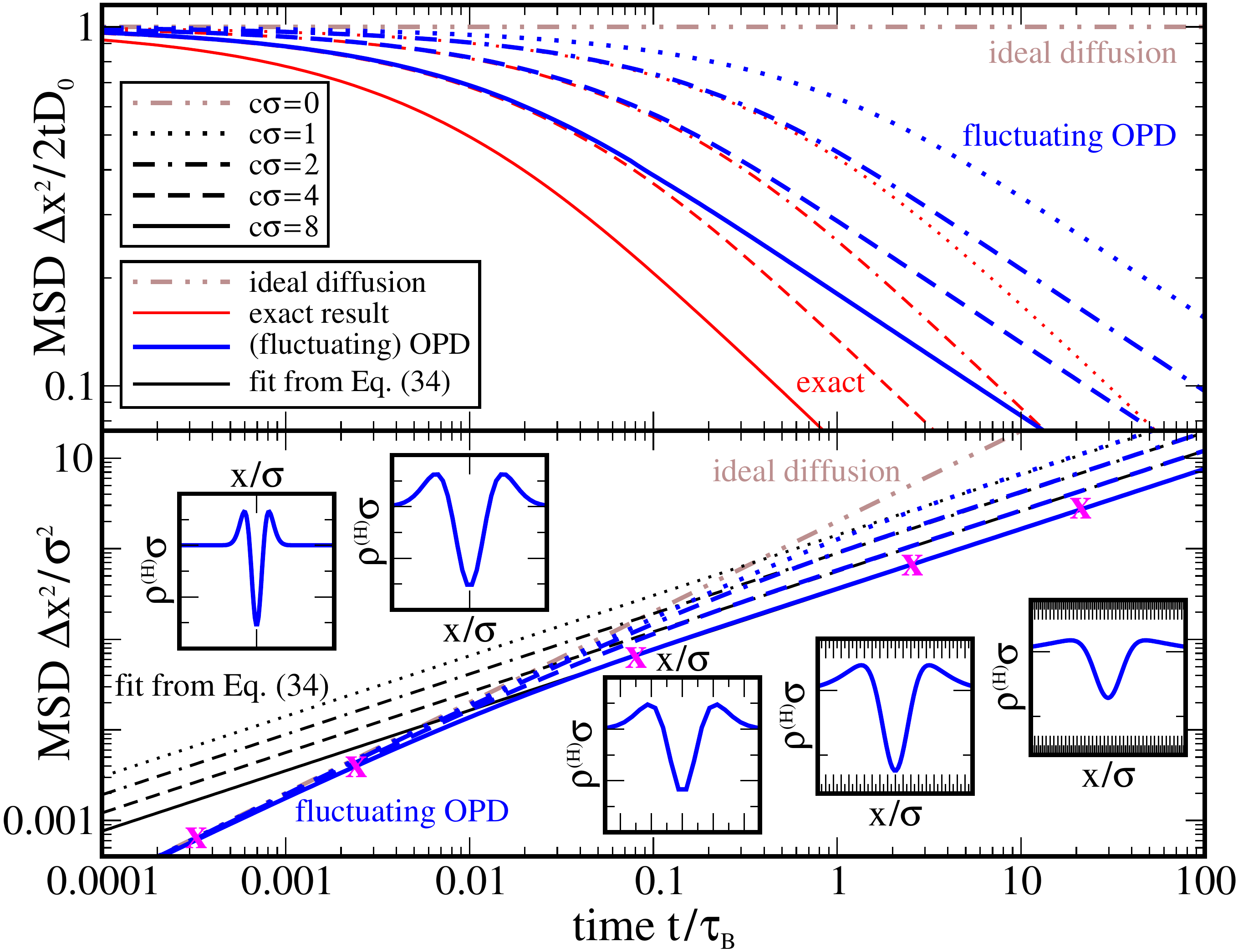} 
\caption{MSD of a tagged point particle {in the limit $N_\alpha\!\rightarrow\!\infty$ using two different normalizations. 
Here, we consider the results from fluctuating OPD, which is equivalent to canonical OPD in this limit. A figure detailing the equivalence is attached to this manuscript as Fig.~\ref{fig_MSDlimCL}.
Considering different concentrations $c$ of host particles (legend with line styles),
we compare the OPD data (labels and legend with color code) to exact results and the diffusion of an ideal particle (no host particles)
and illustrate the match with Eq.~\eqref{eq_MSDOPD} in the long-time limit.
The insets are} as in Fig.~\ref{fig_MSDfin} with $\rho^{(\HH)}(\pm\infty,t)=c$ and additional data points for $t\!\approx\!0.082\tau_\text{B}$ and $t\!\approx\!21.0\tau_\text{B}$).
\label{fig_MSDlim}
}\end{figure}

Here, the dynamics with fluctuating particle numbers are dramatically different.
While the initial decrease of the MSD for each given concentration is similar to that in canonical OPD,
the tagged particle obeys the ideal diffusion law the long-time limit, which is approached via a transient superdiffusive regime.
Such an unphysical behavior has been previously reported {for DDFT} in higher dimensions \cite{hopkinsDTP2010,stopper2015PRE,stopper2015JCP,stopper_bulk}.
 It was declared as an intrinsic drawback of {the theory} that the predicted distinct part of the van Hove function 
(which corresponds the total distribution $\rho^{(\HH)}(x,t)=\rho^{(\LL)}(x,t)+\rho^{(\RR)}(x,t)$ of host particles in our case)
tends to a constant bulk profile \cite{stopper_bulk}.
Indeed, regarding the insets of Fig.~\ref{fig_MSDfin}, we make here a similar observation for fluctuating OPD, also reflecting the distinct equilibrium distributions in the underlying ensembles from Fig.~\ref{fig_EQUILIBRIUM}.
The shape of $\rho^{(\HH)}$ changes from bimodal to unimodal when the tagged particle enters the superdiffusive regime and eventually tends to a Gaussian distribution of the same width as $\rho^{(\TT)}$.
In case of canonical OPD, the host-particle distribution remains bimodal for all times and thus never resembles that of the tagged particle.

Interestingly, for a fixed concentration, all results of fluctuating OPD collapse on the same curve until the inflection point is reached {\cite{SI}}.
This happens at later times for larger (average) numbers of host particles, since the distribution of host particles does not immediately decay towards the sides and remains bimodal for a longer time.
This implies that, if $N_\alpha$ is also taken to infinity, the system will never enter the unphysical subdiffusive regime, which we detail next.

\subsubsection{Infinite system with infinitely many particles \label{sec_SFDlim}}

As a final step, we {study} 
the limit $l\rightarrow\infty$ and $N_\alpha\rightarrow\infty$, such that the concentration $c$ remains finite.
{Upon} subsequently increasing the number of host particles, the qualitative behavior of canonical OPD begins to slightly deviate from the exact curves, which {appear to} converge more rapidly.
Hence, the dynamical slow-down in OPD gets more and more delayed, but eventually a universal subdiffusive long-time limit is approached. 
Intriguingly, the OPD with fluctuating particle numbers converges to the same curve despite the completely opposite behavior for finitely many particles {\cite{SI}}.
As elaborated in Sec.~\ref{sec_SFDfin}, this is reflected by a bimodal distribution of host particles, which maintains the structural information in the distinct part of the van Hove function {for all times}.

In fact, taking the limit is {computationally} much easier in the fluctuating OPD approach based on the gcg ensemble, since it just amounts to extending the system at fixed concentration (or chemical potential), 
while the particle numbers $N_\alpha$ explicitly enter the calculation in canonical OPD, contrast Eq.~\eqref{eq_config} with Eq.~\eqref{eq_gconfig} and compare appendix~\ref{app_ideal}.
Hence, we discuss in the following the limiting curves of fluctuating OPD, {shown in Fig.~\ref{fig_MSDlim}}, which conveniently represent the generic OPD results in this case.

As {remarked} for finite $N_\alpha$, a decrease in the concentration delays the onset of the subdiffusive behavior, in the exact case and for OPD alike.
At short times it even seems like the deviation due to the adiabatic approximation can be absorbed in a renormalization of the concentration $c$ by a factor of two.
The long-time behavior of OPD is reflected nicely by the analytic formula given by Eq.~\eqref{eq_MSDOPD}.
However, it quantitatively deviates from the exact result with the subdiffusive exponent $1/2$.
The larger exponent $2/3$ is most likely an artifact of the adiabatic approximation, maybe together with considering only three species.
While the clarification of this point requires further studies, we stress here that the current DDFT approach is able to successfully describe subdiffusive long-time behavior without any empirical input.

\section{Conclusions \label{sec_conclusion}}

\paragraph*{Ordered ensembles in statistical mechanics.}
The goal of this work is to better understand under which conditions
 a variational theory based on ensemble-averaged one-body fields, like dynamical density functional theory (DDFT) - and also the more general power functional theory, are able to describe (dynamical) caging scenarios. 
As a proof of principle, this problem is boiled down to a clean study of one-dimensional single files.
For all purposes of describing an ergodic equilibrium system there exists an exact equilibrium density functional in one dimension. 
To properly tackle the nonergodic problem of particle order, however, the whole underlying statistical mechanics,
 which defines the notion of an exact density functional, needs to be revisited.
This is done here by introducing an asymmetric ordering potential within three different statistical ensembles in a way 
such that the expectation value of the appropriate density operators gives rise to ordered
equilibrium density profiles of different species.

\paragraph*{Ordered ensembles in density functional theory.}
We argue that for more complex problems than those considered here, the statistical integrals can be treated by employing DFT methods on the level of conditional densities, assuming fixed positions of the tagged particle.
Given its similarity to Percus' test-particle theory, which can be used to get more accurate results out of approximate (mean-field) functionals~\cite{archerevans2017},
our conditional DFT approach might also be beneficial for problems in which one is not interested in individual particles.
One example might be the unphysical crystallization predicted by an effective mean-field functional mimicking elastic interactions \cite{cremer2017}.
Also our dynamical results from {Sec.~\ref{sec_compareTOTAL} and appendix~\ref{sec_compareTOTALasym}} might point to such an improvement, as further discussed below.
Finally, we also discuss the statistical mechanics in the ggg ensemble, treating three ordered species in a grand-canonical fashion.
This provides a crucial first step towards a full DFT treatment of the ordering problem, which is presently still missing.
 A promising candidate for such an approach, which requires an explicit implementation of the asymmetric ordering potential, could be the fundamental-measure functional for nonadditive mixtures \cite{schmidtNAM}.

 \paragraph*{Effect of the adiabatic approximation.} 
 One of the key motivations of our work is to identify a dynamical theory, which in one dimension overcomes the unphysical mixing of neighboring particles over time and thus outputs qualitatively correct tagged-particle profiles.
  Owing to the recent progress since the introduction of the superadiabatic power functional theory \cite{PFT1}, the adiabatic approximation employed for this purpose can by now be perceived as a controlled approximation.
  Hence, the discussed quantitative deviations between this order-preserving dynamics (OPD) and the exact reference results are not a drawback of the underlying equilibrium framework developed here.
 Our most intriguing observation is the different behavior on the level of the total density between OPD and particle-conserving dynamics (PCD) without implicit particle order, despite equal initial conditions and long-time behavior.
 The adiabatic approximation thus seems to act in a different way on different equilibrium functionals.

 \paragraph*{Relation to power functional theory.}
 We stress here that within power functional theory the superadiabatic forces are distinguished from adiabatic ones by their additional functional dependence on the one-body current \cite{PFT1}.
 In this respect, {it is interesting that OPD as an adiabatic theory yields}
 exact results at the system walls, {where} the total contact current vanishes due to the zero-flux boundary condition. 
 {We thus speculate that
 the remaining superadiabatic contributions to the respective power functional might have a simpler structure if OPD with the particle order already imprinted is used as the starting point to tackle dynamical caging effects.} 
To resolve this issue in full detail, it will be enlightening to {construct and add} 
the explicit superadiabatic terms {to both PCD and OPD} \cite{pftapp1}
{or to perform a more careful numerical analysis of the relevant forces \cite{customflow,PFTxPRLseparation}}.
In particular, the exponent in single-file diffusion (SFD) could be revisited within the dynamical test-particle limit of power functional theory~\cite{PFTtestparticle},
 where the superadiabatic van Hove current is expected to counteract the overestimated increase of the MSD \cite{schindler2016}.
 
 \paragraph*{Long-time dynamical behavior.}
 Even within the adiabatic approximation, OPD yields some striking results on the mean-square displacement (MSD), which are not obvious for a variational theory.
While all previous DDFT approaches revert to the standard diffusive behavior of an ideal gas after long times,
 OPD can predict both density-dependent self diffusion and subdiffusive behavior.
In particular, the latter scenario illustrates the dynamical caging effects,
responsible for the distinct part of the van Hove function to never decay to the constant bulk value.
This demonstrates the striking importance to incorporate the particle order 
directly into the underlying equilibrium theory.
In the one-dimensional context of our study this is conveniently achieved by an asymmetric ordering pair potential.

\paragraph*{Outlook.} 

The findings of our study along with Refs.~\cite{schindlerproject,reinhardtbrader2012} demonstrate that, even if the system is ergodic,
a DDFT based on symmetric pair potentials provides an unphysical mathematical shortcut to a tagged particle, which amounts to randomly swapping position with its host particles.
This significantly speeds up the dynamics beyond the effect of the adiabatic approximation.
It thus remains a challenging task for future research to properly address the caging scenario within DDFT in higher dimensions.
 A first step in this direction could consist of employing a DDFT based on Eq.~\eqref{eq_rhoWcanr} and Eq.~\eqref{eq_rhoIcanGENr}.
 Going beyond this extended test-particle approach would, however, require the definition of a many-body caging potential, as an appropriate generalization of Eq.~\eqref{eq_WNalpha}. 
Alternatively, some simple nonergodic quasi-one-dimensional situations should be addressed using conditional DFT and OPD 
to learn more about the approximations made in the density functionals in higher dimensions.
 A more direct application of our methods would be to treat the tagged particle as an adiabatic piston separating two subsystems in any spatial dimension \cite{foulaadvand2013,lieb1999,caprini2017,khalil2019}.
Establishing a closer connection between DDFT and alternative solution methods of the diffusion equation through boundary conditions \cite{bruna2012a,bruna2012b,bruna2014} could also prove fruitful in this context.
Finally, regarding the one-dimensional case, OPD can be readily applied to some more general diffusion problems including particles with soft interactions \cite{ambjornsson2008b,nelissen2007}, external forcing \cite{taloni2006, ryabov2011}, 
or SFD of models for active particles~\cite{doussal2019}.

\appendix

\section{Some background on density functional theory (DFT) \label{app_DDFT}}

In this appendix we give a brief account of density functional theory (DFT) \cite{evans79,bob92} 
and describe how the calculations from the main text can be redone by explicitly using DFT methods.

\subsection{DFT in a nutshell}

Classical DFT is a versatile tool to determine the equilibrium one-body density profile $\rho_0(\bvec{r})$
of a fluid of interacting particles in any given external potential $\Vext(\bvec{r})$ acting on the centers of each particle.
It relies on the fundamental principle that, from any given $\rho_0(\bvec{r})$ and pair interaction $u(r)$, the external potential can be uniquely inferred \cite{Mermin}.
With this background, it can be shown that for a given a density functional $\Omega[\rho(\bvec{r})]$ the equilibrium density profile follows from the variational Euler-Lagrange equation
\begin{align}
 \frac{\delta\Omega[\rho(\bvec{r}')]}{\delta\rho(\bvec{r})}=0\,,
 \label{eq_ELG}
\end{align}
where $\Omega[\rho_0(\bvec{r})]$ equals the grand potential $\Omega$ of the system and $\Omega[\rho]>\Omega$ if $\rho\neq\rho_0$.
Hence, if $\Omega[\rho(\bvec{r})]$ is known, all structural and thermodynamical properties of a system can be inferred.
The explicit form of a density functional is discussed in the following for the case of a conditional system
 required to reproduce the definitions from Sec.~\ref{sec_gc} in the gcg ensemble. 

\subsection{Conditional DFT in one dimension \label{app_condDFT}}

Assuming that, in a one-dimensional system, the tagged particle is fixed at position $x_0$
we can formulate a conditional DFT in terms of the conditional density $\varrho^{(\alpha)}(x|x_0)$ to find a host particle at position $x\leq x_0$ for $\alpha={\LL}$
or $x\geq x_0$ for $\alpha={\RR}$.
In this case, all functionals receive an explicit dependence on the position of the tagged particle and carry a species label $\alpha$, while there is otherwise no difference to standard DFT for a single species.

The appropriate conditional density functional 
\begin{align} 
\varOmega^{(\alpha)}[\varrho^{(\alpha)}](x_0)
&=\mathcal{F}_\text{id}+\mathcal{F}_\text{ex}+\varOmega_\text{ext}^{(\alpha)}
\label{eq_DFrho2}
\end{align}
can be separated into three terms.
The ideal part of the intrinsic Helmholtz free energy functional is exactly given by the general form
\begin{align} 
 \beta \mathcal{F}_\text{id}[\varrho^{(\alpha)}](x_0)=\int\upd x\,\varrho^{(\alpha)}(x|x_0)\, \left(\ln(\Lambda\varrho^{(\alpha)}(x|x_0))-1\right),
\end{align}
which is the same in any spatial dimension.
In one dimension, there is also an exact expression for the excess free energy functional \cite{percus,percus2}
\begin{align} 
 \mathcal{F}_\text{ex}[\varrho^{(\alpha)}](x_0)=-\int\upd x\, n_0(x|x_0)\ln\left(1-n_1(x|x_0)\right),
\end{align}
indicating the contribution due to interparticle pair interactions, cf. Eq.~\eqref{eq_u}.
This hard-core potential is split into the two weight functions $\omega^{(0)}(x)=\frac{1}{2}\left( \delta(R-x) + \delta(R+x) \right)$
and $\omega^{(1)}(x)=\Theta(R-|x|)$ with the Heaviside step function $\Theta(x)$,
used to construct the (conditional) weighted densities
\begin{align} 
n_i(x|x_0)= \int\upd x'\,\varrho^{(\alpha)}(x'|x_0)\,\omega^{(i)}(x-x')\,,
\label{eq_DFrhoEWD}
\end{align}
where $i\in\{0,1\}$.
Finally, the extrinsic part 
\begin{align}
\!\!\!\!\!\!&\varOmega_\text{ext}^{(\alpha)}[\varrho^{(\alpha)}](x_0)\cr&\ \ = \int\upd x\,\varrho^{(\alpha)}(x|x_0)\left( w_\alpha(x-x_0)+\Vext^{(\alpha)}(x)-\mu_\alpha \right)\ \ \ \ \ \ 
\end{align}
of the functional contains the standard contributions of external potential $\Vext^{(\alpha)}(x)$ and chemical potential $\mu_\alpha$
together with the additional order-preserving potential $w_\alpha(x-x_0)$ from Eq.~\eqref{eq_w}, which formally takes the role of an additional external potential, as discussed in Sec.~\ref{sec_c}.
 Note, however, that this additional term is actually part of the intrinsic free energy when regarding the system as a whole, since it represents an interparticle interaction.
Therefore, in the context of DDFT, there is no explicit contribution arising from $w_\alpha$ in Eq.~\eqref{eq_DDFTgc} of the main text.

Solution of the Euler-Lagrange equation, Eq.~\eqref{eq_ELG}, with the conditional functional from Eq.~\eqref{eq_DFrho2}
yields the conditional equilibrium densities $\varrho_0^{(\alpha)}(x|x_0)$ and conditional grand-canonical partition functions 
\begin{align} 
 \varXi^{(\alpha)}(x_0)=\exp\left(-\beta\varOmega^{(\alpha)}(x_0)\right)\,,
\end{align}
where $\varOmega^{(\alpha)}(x_0):=\varOmega^{(\alpha)}[\varrho_0^{(\alpha)}](x_0)$ are the conditional grand potentials.
With the help of these two quantities and Eq.~\eqref{eq_gconfig} of the main text,
the ordered density profiles $\rho_\text{gcg}^{(\TT)}(x)$ and $\rho_\text{gcg}^{(\alpha)}(x)$ in the gcg ensemble follow from Eq.~\eqref{eq_rhoIgcT} and Eq.~\eqref{eq_rhoIgcGEN} of the main text, respectively.
Note that in the main text and in the following the index 0 denoting equilibrium is dropped for convenience.

\subsection{Canonical inversion method}

The conditional density functional from Eq.~\eqref{eq_DFrho2} can also be used to calculate the densities in an ordered canonical system.
To this end, the inversion method introduced in Ref.~\cite{delasheras2014} can be directly applied,
without requiring a generalization to mixtures as in Ref.~\cite{schindlerproject}.

 Analogously to Eq.~\eqref{eq_Xigcg} of the main text, the conditional grand canonical partition functions
 \begin{align}
 \varXi^{(\alpha)}(x_0)&=\sum_{N_\alpha=0}^\infty e^{\beta\mu_\alpha N_\alpha}\mathcal{\CPF}^{(\alpha)}_{N_\alpha}(x_0) \,,
\label{eq_configGC}
\end{align}
 can be obtained from an infinite sum over the canonical ones.
The same can be done for the conditional densities
\begin{align}
 \varrho^{(\alpha)}(x|x_0)=\sum_{N_\alpha=1}^{\infty}p_{N_\alpha}(x_0) \varrho^{(\alpha)}_N(x|x_0)
 \label{eq_CIM2}
\end{align}
with the position-dependent probability
\begin{align}
 p_{N_\alpha}(x_0)=e^{\beta\mu_\alpha N_\alpha}\frac{\mathcal{\CPF}^{(\alpha)}_{N_\alpha}(x_0)}{\varXi^{(\alpha)}(x_0)}
\end{align}
 to find $N_\alpha$ particles in a subsystem for the given chemical potential $\mu_\alpha$.
 
 In a confined system, the sums in Eqs.~\eqref{eq_configGC} and~\eqref{eq_CIM2} can be truncated at the maximal particle numbers $M_\alpha$,
 which, in each case, yields a system of $M_\alpha$ equations for the $M_\alpha$ conditional canonical partition functions and densities (removing the trivial case of an empty system).
  Having extracted these quantities for the desired $N_\alpha$, we can calculate $\rho_N^{(\TT)}(x)$ and $\rho_N^{(\alpha)}(x)$ according Eq.~\eqref{eq_rhoIcanGEN} of the main text.
    As all ingredients to the one-dimensional DFT are exact, this procedure is completely equivalent to defining the conditional quantities from statistical integrals.

\section{Analytical results for point particles \label{app_ideal}}

Here we state and discuss the ordered equilibrium profiles of point particles in the different ensembles, shown in Fig.~\ref{fig_EQUILIBRIUM},
To this end, we use the definitions from Sec.~\ref{sec_ensembles} with vanishing particle radius $R=0$.
Hence the pair potential $u(|x|)$ from Eq.~\eqref{eq_u} becomes constantly zero, while this is not the case for the ordering potential $w(x)$ from Eq.~\eqref{eq_w}.
The external potentials $\Vext^{(\nu)}(x)$ acting on all species $\nu$ alike model hard walls at $x=0$ and $x=l$, 
 such that the Boltzmann factor $e^{-\beta\Vext^{(\nu)}(x)}=\Theta(x)\Theta(l-x)$ sets the boundaries of the integrals.
 This factor is common to all of the following densities and conditional partition functions, such that we drop it for notational convenience.

\subsection{Ordered canonical ensemble \label{app_can}}

As defined in Eq.~\eqref{eq_config}, the conditional canonical partition function
\begin{align}
\mathcal{\CPF}_{N-1}(x_0)=\frac{x_0^{N_{\LL}}}{N_{\LL}!\Lambda^{N_{\LL}}}\,\frac{(l-x_0)^{N_{\RR}}}{N_{\RR}!\Lambda^{N_{\RR}}}\label{eq_QnID}
\end{align}
is a product of the contributions $\mathcal{\CPF}^{(\alpha)}_{N_{\LL}}(x_0)$ from the two subsystems left and right to the tagged particle.
Integration gives the total canonical partition function $\CPF_N=l^N/(N!\,\Lambda^N)$, identical to that of $N$ ideal indistinguishable particles (without explicit order).

As the conditional densities $\varrho_N^{({\LL})}(x|x_0)=N_{\LL}/x_0$ and $\varrho_N^{({\RR})}(x|x_0)=N_{\RR}/(l-x_0)$ of the two host-particle species are simply constants (as a function of $x$) between the wall and the tagged particle,
we find from Eq.~\eqref{eq_rhoIcanGEN},
\begin{align}
\rho^{(\TT)}_N(x) &= \frac{N!}{l^NN_{\LL}!\,N_{\RR}!}\,x^{N_{\LL}}(l-x)^{N_{\RR}}\,,\cr
  \rho_N^{(\LL)}(x)&=\sum_{n=\,0}^{N_{\LL}-1} \left.\rho_N^{(\TT)}(x)\right|_{N_{\LL}=n,N_{\RR}=N-1-n} \,,\cr
  \rho_N^{(\RR)}(x)&=\sum_{n=\,N_{\LL}+1}^{N-1} \left.\rho_N^{(\TT)}(x)\right|_{N_{\LL}=n,N_{\RR}=N-1-n}\,.\label{eq_rhoAPPcan}
 \end{align}
  From these series representations of the distributions of the host particles,  
  the relation in Eq.~\eqref{eq_sumrhoc} can be directly inferred.
  Closed expressions can only be given for explicit choices of $N_{\LL}$ and $N_{\RR}$.

\subsection{Ordered gcg ensemble}
 
 In the gcg ensemble with the tagged particle treated as a canonical species and the numbers of host particles allowed to fluctuate, we find
 the partition function
 \begin{align}
\Xi_\text{gcg}=\frac{1}{\Lambda}\frac{e^{lz_{\LL}}-e^{lz_{\RR}}}{z_{\LL}-z_{\RR}}\ \ \stackrel{\mu_{\HH}}{\longrightarrow} \ \ \frac{l}{\Lambda}\,e^{lz_{\HH}}\,,
\label{eq_MXiID}
\end{align}
introducing the activity
 \begin{align}
  z_\nu=\frac{e^{\beta\mu_\nu}}{\Lambda}
 \end{align}
of species $\nu$, reflecting the bulk density of an ideal gas.
 The second result indicated by the arrow denotes the symmetric situation, considered here, of equal chemical potentials $\mu_{\HH}:=\mu_{\LL}=\mu_{\RR}$ (and thus average particle numbers) of the two host species.
 The densities read
 \begin{align}
 \rho^{(\TT)}_\text{gcg}(x)&=\frac{e^{xz_{\LL}}\,e^{(l-x)z_{\RR}}}{\Lambda\, \Xi_\text{gcg}}\ \ \stackrel{\mu_{\HH}}{\longrightarrow} \ \ \frac{1}{l}\,,\cr
 \rho^{(\LL)}_\text{gcg}(x)&=\frac{e^{lz_{\LL}}-e^{(l-x)z_{\RR}+xz_{\LL}}}{\Lambda\,(z_{\LL}-z_{\RR})\, \Xi_\text{gcg}}\ \ \stackrel{\mu_{\HH}}{\longrightarrow} \ \ \frac{l-x}{l}\,z_{\HH}\,,\cr
 \rho^{(\RR)}_\text{gcg}(x)&=\frac{e^{(l-x)z_{\RR}+xz_{\LL}}-e^{lz_{\RR}}}{\Lambda\,(z_{\LL}-z_{\RR})\, \Xi_\text{gcg}}\ \ \stackrel{\mu_{\HH}}{\longrightarrow} \ \ \frac{x}{l}\,z_{\HH}\,.\label{eq_rhoAPPgcg}
\end{align}
  To illustrate the equivalence of the different methods, we discuss below three ways to obtain these universal formulas.

 The first method is to take the canonical results from Sec.~\ref{app_can} and employ Eqs.~\eqref{eq_Xigcg} and~\eqref{eq_rhoGCG} of the main text.
 This can be conveniently done by properly rearranging the sums.
 The second method reverts to the conditional results in the gcg ensemble. 
 Summation of the two factors in Eq.~\eqref{eq_QnID} according to Eq.~\eqref{eq_configGC} yields
   \begin{align}
 \varXi(x_0)=\varXi^{(\LL)}(x_0)\,\varXi^{(\RR)}(x_0)=e^{x_0z_{\LL}}\,e^{(l-x_0)z_{\RR}}. \label{eq_XiID}
\end{align}
 As in the canonical case, the conditional densities $\varrho_\text{gcg}^{(\alpha)}(x|x_0)=z_{\alpha}$, obtained from Eq.~\eqref{eq_CIM2}, are simply the constant ideal-gas results between the wall and the tagged particle. 
 It is then easy to see that Eqs.~\eqref{eq_gconfig}, \eqref{eq_rhoIgcT} and~\eqref{eq_rhoIgcGEN} of the main text give the same results as from the first method,
 where the integral runs from $x_0$ to $l$ for $\alpha=\LL$ and from $0$ to $x_0$ for $\alpha=\RR$.
 Finally, using the conditional DFT from Sec.~\ref{app_condDFT}, the Euler-Lagrange equations corresponding to the functionals from Eq.~\eqref{eq_DFrho2} read
 $\beta^{-1}\ln(\Lambda\varrho_\text{gcg}^{(\alpha)})-\mu_\alpha=0$
 in the region between the wall and the tagged particles, as $\mathcal{F}_\text{ex}=0$ for ideal point particles.
  The other regions require $\varrho_\text{gcg}^{(\alpha)}=0$, since either $w_\alpha(x-x_0)$ or $\Vext^{(\alpha)}(x)$ is infinite.
 We easily see that $\varrho_\text{gcg}^{(\alpha)}(x|x_0)=z_{\alpha}$ solve these equations, as expected for an ideal gas.
 Inserting these back into the functionals gives Eq.~\eqref{eq_XiID}, since $\varXi^{(\alpha)}(x_0)=\exp(-\beta\varOmega^{(\alpha)}(x_0))$ with
 \begin{align} 
\beta\varOmega^{(\LL)}(x_0)\!=\!\!\int_0^{x_0}\upd x\,z_\LL\, \left(\ln(e^{\beta\mu_-})-1-\beta\mu_- \right)=-x_0z_\LL
\end{align}
 and the same for $\varOmega^{(\RR)}=-(l-x_0)\,z_\RR$.
 The remaining steps are the same as for the second method.

Comparing the densities from Eq.~\eqref{eq_rhoAPPcan} and Eq.~\eqref{eq_rhoAPPgcg}, we see that increasing the system at fixed concentration $c=2N_\alpha/l$ of host particles
simply amounts to a change of $l$ at constant chemical potentials in the gcg ensemble,
while, in the canonical case, the particle numbers $N_\alpha$ need to be increased proportionally.
As mentioned in Sec.~\ref{sec_SFDlim}, the calculation of $\rho^{(\nu)}_N$ is thus computationally more difficult than that of $\rho^{(\nu)}_\text{gcg}$ for large systems.

\subsection{Ordered ggg ensemble}

The densities in the ggg ensemble, which treats all species grand-canonically,
can only be found by the first method discussed for the gcg ensemble, that by Eqs.~\eqref{eq_Xiggg} and~\eqref{eq_rhoGGG} of the main text.
Restricting ourselves to the case $\mu_{\HH}=\mu_{\LL}=\mu_{\RR}$, such that $\rho^{(\TT)}_\text{ggg}(x)=\rho^{(\TT)}_\text{ggg}(l-x)$ is generally symmetric and $\rho^{(\RR)}_\text{ggg}(x)=\rho^{(\LL)}_\text{ggg}(l-x)$, the results are
 \begin{align}
 \rho^{(\TT)}_\text{ggg}(x)&=\frac{z_{\HH}^2z_{\TT}e^{lz_{\HH}}+z_{\TT}^3e^{lz_{\TT}}}{(z_{\HH}-z_{\TT})^2 \Xi_\text{ggg}}\cr
 &\ \ \ \ - \frac{z_{\HH}z_{\TT}^2(e^{xz_{\HH}+(l-x)z_{\TT}}+e^{xz_{\TT}+(l-x)z_{\HH}})}{(z_{\HH}-z_{\TT})^2 \Xi_\text{ggg}}\,,\ \ \ \ \cr
 \rho^{(\LL)}_\text{ggg}(x)&=\frac{((l-x)z_{\HH}(z_{\HH}-z_{\TT})-z_{\TT})z_{\HH}z_{\TT}e^{lz_{\HH}}}{(z_{\HH}-z_{\TT})^2 \Xi_\text{ggg}}\cr
 &\ \ \ \ + \frac{z_{\HH}z_{\TT}^2e^{xz_{\HH}+(l-x)z_{\TT}}}{(z_{\HH}-z_{\TT})^2 \Xi_\text{ggg}}\,,\cr
 \rho^{(\RR)}_\text{ggg}(x)&=\frac{(xz_{\HH}(z_{\HH}-z_{\TT})-z_{\TT})z_{\HH}z_{\TT}e^{lz_{\HH}}}{(z_{\HH}-z_{\TT})^2 \Xi_\text{ggg}}\cr
 &\ \ \ \ + \frac{z_{\HH}z_{\TT}^2e^{(l-x)z_{\HH}+xz_{\TT}}}{(z_{\HH}-z_{\TT})^2 \Xi_\text{ggg}}\label{eq_rhoAPPggg}
\end{align}
 with the partition function
\begin{align}
\!\!\!\!\!\!\Xi_\text{ggg}=\frac{\left((lz_{\HH}^2+2z_{\HH})(z_{\HH}-z_{\TT})-z_\HH^2\right)e^{lz_{\HH}}+z_{\TT}^2e^{lz_{\TT}}}{(z_{\HH}-z_{\TT})^2}\,.\!\!\!\!
\label{eq_MXiIDggg}
\end{align}
Further equating all chemical potentials $\mu=\mu_{\TT}=\mu_{\HH}$ (and $z=z_{\TT}=z_{\HH}$) yields
 \begin{align}
 \rho^{(\TT)}_\text{ggg}(x)&=\frac{2z(xz+1)((l-x)z+1)}{(lz+2)^2-2}\,,\cr
 \rho^{(\LL)}_\text{ggg}(x)&=\frac{z^2(l-x)((l-x)z+2)}{(lz+2)^2-2}\,,\cr
 \rho^{(\RR)}_\text{ggg}(x)&=\frac{z^2x(xz+2)}{(lz+2)^2-2}\,\label{eq_rhoAPPgggMU}
\end{align}
and
\begin{align}
\Xi_\text{ggg}=\frac{((lz+2)^2-2)e^{lz}}{2}\,.
\label{eq_MXiIDgggMU}
\end{align}

\subsection{Comparison of the ensembles in certain limits}

The two ordered ensembles with fluctuating particle numbers introduced in this work are, in general, different from the
generic grand canonical ensemble, whose well-known partition function 
\begin{align}
\Xi=e^{lz}
\label{eq_XiIDexplicit}
\end{align}
follows from Eq.~\eqref{eq_XiGC} of the main text. For a true mixture of three species without enforced particle order the partition function is just a product of different factors $e^{lz_\nu}$.
Apparently, none of these functions is recovered when equating all chemical potentials in the ordered gcg or ggg ensembles, as in Eq.~\eqref{eq_MXiID} and Eq.~\eqref{eq_MXiIDgggMU}, respectively.
The only way to recover the grand-canonical single-species result, Eq.~\eqref{eq_XiIDexplicit}, from Eq.~\eqref{eq_MXiIDggg} 
in the gcg ensemble is by setting the chemical potentials $\mu_\HH$ of the host particles (or in general of two arbitrary species) to minus infinity.
Doing the same to Eq.~\eqref{eq_MXiID} in the gcg ensemble results int the canonical partition function $\CPF_1=l/\Lambda$ for a single particle with $N_\TT=1$.

The more interesting question, addressed in Fig.~\ref{fig_EQUILIBRIUM},
concerns the relations between the considered ordered ensembles.
To this end, we explicitly calculate the limiting behavior of the density profiles $\rho^{(\nu)}_\text{ggg}$,
given by Eq.~\eqref{eq_rhoAPPggg} or Eq.~\eqref{eq_rhoAPPgggMU}, in the gcg ensemble in four different cases.
Note that the average numbers $N_\nu=\int_0^l\upd x\,\rho^{(\nu)}_\text{ggg}(x)$ of particles in each species are monotonously increasing functions $N_\nu(z_\HH,z_\TT)$ of all activities (or chemical potentials),
where $N_\LL(z_\HH,z_\TT)=N_\RR(z_\HH,z_\TT)\neq N_\TT(z_\HH,z_\TT)$.

We first realize that for large and equal particle numbers in each species, the activities $z_\HH\simeq z_\TT$ approach each other.
Hence, we can take from Eq.~\eqref{eq_rhoAPPgggMU} the many-particle limit
 \begin{align}
 \rho^{(\TT)}_\text{ggg}(x)&\stackrel{\Lambda z\gg1}{=}\frac{2x(l-x)z}{l^2}\,,\cr
 \rho^{(\LL)}_\text{ggg}(x)&\stackrel{\Lambda z\gg1}{=}\frac{2(l-x)^2z}{l^2}\,,\cr
 \rho^{(\RR)}_\text{ggg}(x)&\stackrel{\Lambda z\gg1}{=}\frac{2x^2z}{l^2}\,
\end{align}
of the densities, assuming that $N_\LL=N_\TT=N_\RR$.
These functions have the same form as the density profiles in the canonical ensemble, Eq.~\eqref{eq_rhoAPPcan},
when choosing all particle numbers $N_\nu=1$ equal to one.

In the second case, we assume the opposite limit of very small particle numbers through a Taylor expansion in the activities up to second order.
The resulting expressions
 \begin{align}
 \rho^{(\TT)}_\text{ggg}(x)&\stackrel{\Lambda z_\nu\ll1}{=}(1-lz_\HH)z_\TT\,,\cr
 \rho^{(\LL)}_\text{ggg}(x)&\stackrel{\Lambda z_\nu\ll1}{=}(l-x)z_\HH z_\TT\,,\cr
 \rho^{(\RR)}_\text{ggg}(x)&\stackrel{\Lambda z_\nu\ll1}{=}xz_\HH z_\TT\,\label{eq_rhoAPPgggLD}
\end{align}
for the density profiles in this low-density expansion
resemble those in the gcg ensemble, Eq.~\eqref{eq_rhoAPPgcg}.
In this case, integration shows that the particle numbers in the tagged and host species are different. 
Taking into account just the terms in Eq.~\eqref{eq_rhoAPPgggLD} which are linear in the activities, amounts to ignoring the pairwise order-preserving interaction.
In this case, the only nonvanishing contribution is from the constant density $\rho^{(\TT)}_\text{ggg}$ of the tagged particle.

As a slightly more general scenario, we assume that there are much more particles in the host species than in the tagged one,
yielding
 \begin{align}
 \rho^{(\TT)}_\text{ggg}(x)&\stackrel{z_\HH \gg z_\TT}{=}\frac{z_\TT}{1+lz_\HH}\,,\cr
 \rho^{(\LL)}_\text{ggg}(x)&\stackrel{z_\HH \gg z_\TT}{=}\frac{(l-x)z_\HH z_\TT}{1+lz_\HH}\,,\cr
 \rho^{(\RR)}_\text{ggg}(x)&\stackrel{z_\HH \gg z_\TT}{=}\frac{xz_\HH z_\TT}{1+lz_\HH}\,.
\end{align}
Apparently, the above low-density expressions from Eq.~\eqref{eq_rhoAPPgggLD} simply follow from an expansion of the denominator.
Hence, the low-density limit implies $N_\TT\ll N_\HH$, which can be understood by the appearance of the linear term in the activity expansion of $\rho^{(\TT)}_\text{ggg}$.
It is also clear that, again, the density profiles behave as in the gcg ensemble, Eq.~\eqref{eq_rhoAPPgcg}.

Finally, in the opposite limit with much less particles in the host species than in the tagged one, we find
 \begin{align}
 \rho^{(\TT)}_\text{ggg}(x)&\stackrel{z_\HH \ll z_\TT}{=}z_\TT\,,\cr
 \rho^{(\LL)}_\text{ggg}(x)&\stackrel{z_\HH \ll z_\TT}{=}(e^{-xz_\TT}-e^{-lz_\TT})z_\HH \,,\cr
 \rho^{(\RR)}_\text{ggg}(x)&\stackrel{z_\HH \ll z_\TT}{=}(e^{-(l-x)z_\TT}-e^{-lz_\TT})z_\HH \,.\label{eq_rhoAPPgggCGC}
\end{align}
While such an assumption is not of much use for the tagged-particle problem,
it is the only one that shows a clear equivalence between the ggg and canonical ensembles, which we understand as follows.
Choosing the activities $z_\HH$ and $z_\TT$ such that $N_\LL=N_\RR=1$ and letting $N_\TT$ become very large,
the density profiles of the host particles, whose functional form is then given by Eq.~\eqref{eq_rhoAPPgggCGC}, become completely equivalent to those 
\begin{align}
\!\!\!\! \left.\rho^{(\LL)}_N(x)\right|_{N_\LL=1}&=\frac{N(l-x)^{N-1}}{l^N}\stackrel{N \gg1}{=}\frac{N}{l}e^{-(N-1)\frac{x}{l}}\,, \cr
\!\!\!\! \left.\rho^{(\RR)}_N(x)\right|_{N_\RR=1}&=\frac{Nx^{N-1}}{l^N}\stackrel{N \gg1}{=}\frac{N}{l}e^{-(N-1)\frac{l-x}{l}}\,\ \ \ 
\end{align}
in the corresponding canonical case, where the second equalities hold for large $N=N_\TT+2$.
 This scenario can thus be called a proper thermodynamic limit, which, in this sense, does not exist if the tagged species is restricted to holding a single particle.

{ 

\section{Asymmetric initial trapping \label{sec_compareTOTALasym}}

In Fig.~\ref{fig_T3mid} of the main text, 
we discussed the density profiles of $N=3$ hard rods of length $\sigma$ in a slit of length $l=4.9\sigma$,
initialized using a common harmonic trapping potential for each species centered at $x_\text{h}^{(\nu)}=l/2=2.45\sigma$ in the middle of the slit.
In this appendix we elaborate on the observations for a larger system with $l=5.9\sigma$,
where the initial trap is still located at $x_\text{h}^{(\nu)}=2.45\sigma$, which now lies to the left of the center of the box.
Apart from the broken symmetry, this setup allows to determine how the particles move into the initially
nearly depleted region on the right.
Moreover, the density profiles of the two host species may now overlap.

\begin{figure}[t]
\includegraphics[width=0.4425\textwidth] {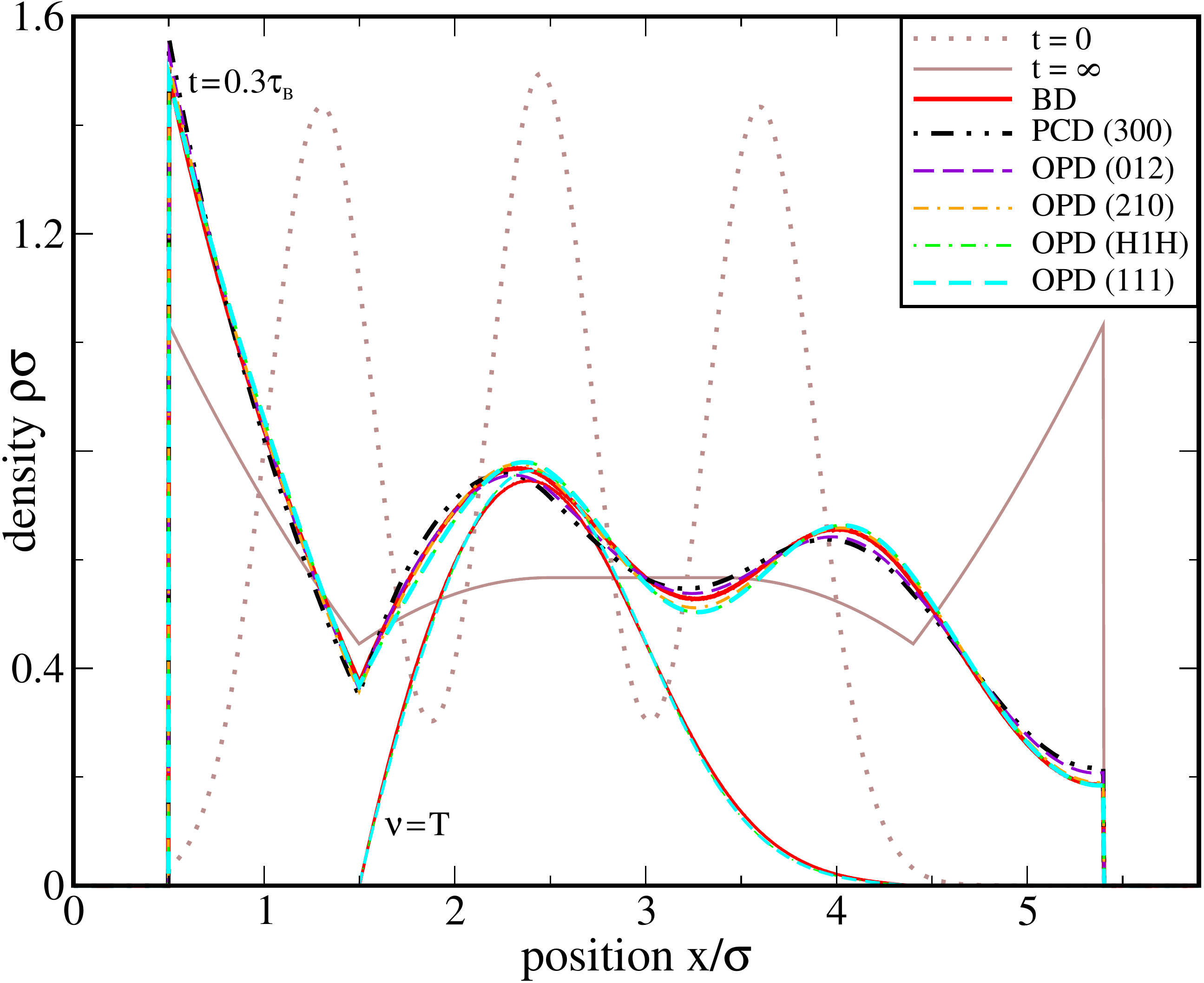} 
\caption{Total density profiles as in Fig.~\ref{fig_T2mid} {of the main text} but for $N\!=\!3$ hard rods and $l\!=\!5.9\sigma$. 
Three additional sets of OPD data are included, each considering two different particles as members of a single species (as labeled). 
The label $(\text{H}1\text{H})$ refers to tagging the central particle but considering the host particles
as members of the same species, as $\rho_N^{(\HH)}=\rho_N^{(\LL)}+\rho_N^{(\RR)}$. 
 A more detailed figure for this setup, including different time steps, is attached to this manuscript as Fig.~\ref{fig_T3largeCL}.
\label{fig_T3large}
}\end{figure}

Figure~\ref{fig_T3large} illustrates that OPD closely captures the spreading of the density to the right.
 In this region, there are only very small differences to the exact BD for both the total density and the density of the central tagged particle.
In particular, the} contact density at the right wall increases in {exactly} the same fashion.
Also the nonmonotonous behavior over time of the contact value at the left wall is perfectly reproduced, {which is consistent with the observations in the main text}.
Between the walls, OPD is again somewhat slower than BD.
In contrast, regarding the decay of the rightmost density peak, PCD
overtakes BD as also observed for the smaller system with $N=2$ in Fig.~\ref{fig_T2mid} {of the main text} but not with $N=3$ in Fig.~\ref{fig_T3mid} {of the main text}.
This points to some density-dependent effects.

With the given asymmetric initial conditions, it also becomes important, which particle is chosen as the tagged particle.
This is examined in Fig.~\ref{fig_T3large} by further comparing the two partial versions of OPD with only two species holding $N_{\TT}=1$ and $N_{\RR}=2$ particles or $N_{\LL}=2$ and $N_{\TT}=1$ particles.
For example, in the latter case, we tag the particle on the right, which has the highest mobility at short times. This results in the initial dynamics being very close to full OPD (except for the region close to the left boundary).
{Finally, choosing again the central particle as the tagged particle, but considering both host particles, left and right, as members of the same species,
results in some deviations compared to the full OPD with three different species, which are, however, barely noticeable.}

\vspace*{-0.1425cm}

\acknowledgments

The authors would like to thank Thomas Schindler, Abhinav Sharma, Michael te Vrugt, Raphael Wittkowski, Suvendu Mandal and Daniel de las Heras for stimulating discussions.
We also acknowledge Thomas Schindler for providing the BD code used in this study and careful proofreading of the manuscript.
 Financial support is provided by the DFG through the SPP 2265, under grant numbers WI 5527/1-1 (R.W.) and LO 418/25-1 (H.L.).
This paper is dedicated to the late Gerhard Findenegg.

\newpage

\begin{figure*}[t]
\includegraphics[width=0.625\textwidth] {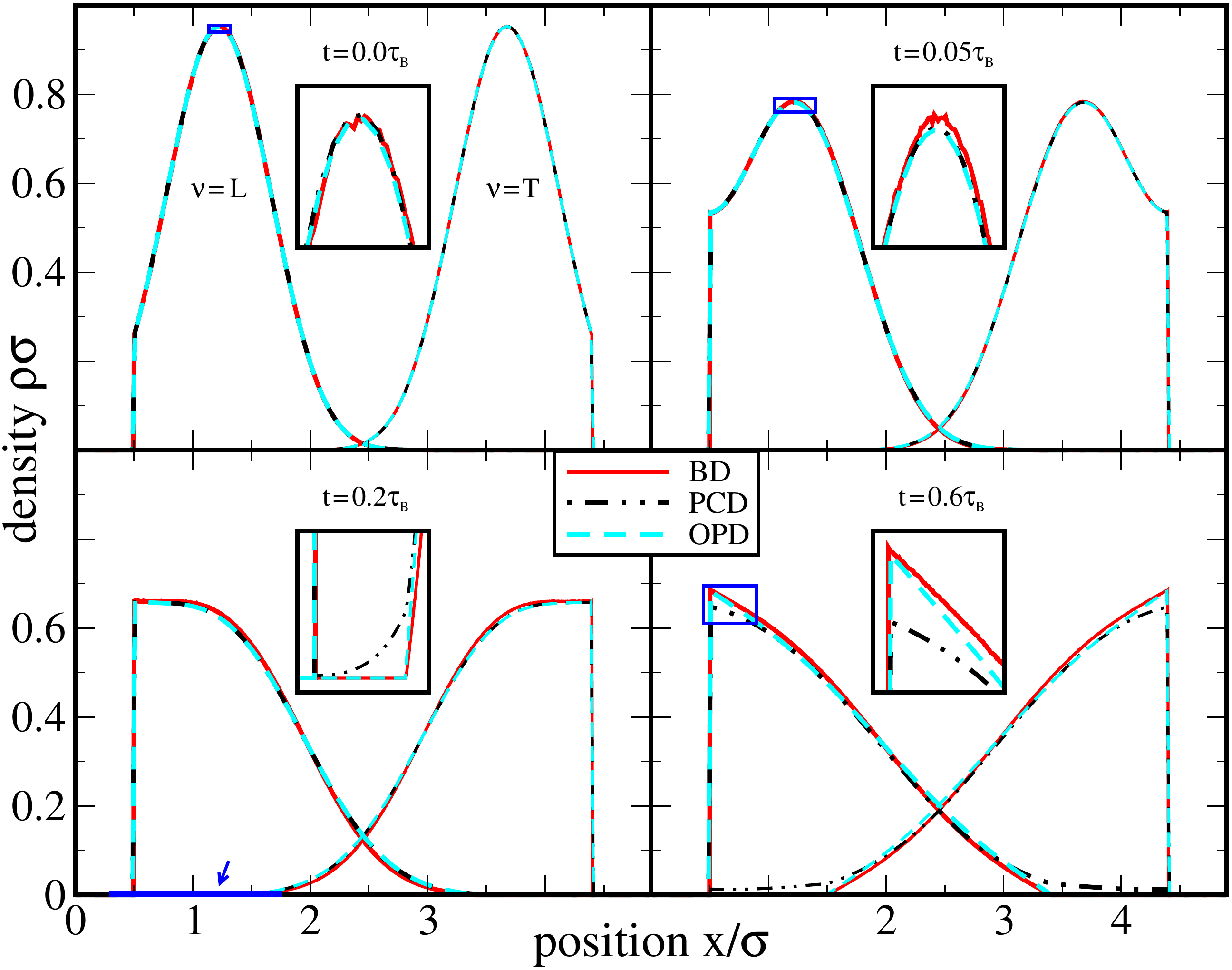} 
\caption{Extension of Fig.~4 from the main manuscript. Shown are the individual density profiles $\rho_N^{(\nu)}(x,t)$ of $N\!=\!2$ hard rods of length $\sigma$ on a line of length $l\!=\!4.9\sigma$ for distinguished initial trapping 
with $k^{(\LL)}\!=\!k^{(\TT)}\!=\!5/(\beta\sigma^2)$, $x_\text{h}^{(\TT)}\!=\!l/4$ and $x_\text{h}^{(\TT)}\!=\!3l/4$ (species $\nu\!=\!{\LL}$ and $\nu\!=\!{\TT}$ as labeled).
We compare BD and PCD results as in Ref.~\cite{schindlerproject} to OPD as labeled, where each subfigure shows the results for a given time.
The insets show closeups of the boxed regions.
\label{fig_T2figCL}
}
\end{figure*}
\begin{figure*}[b]
\includegraphics[width=0.625\textwidth] {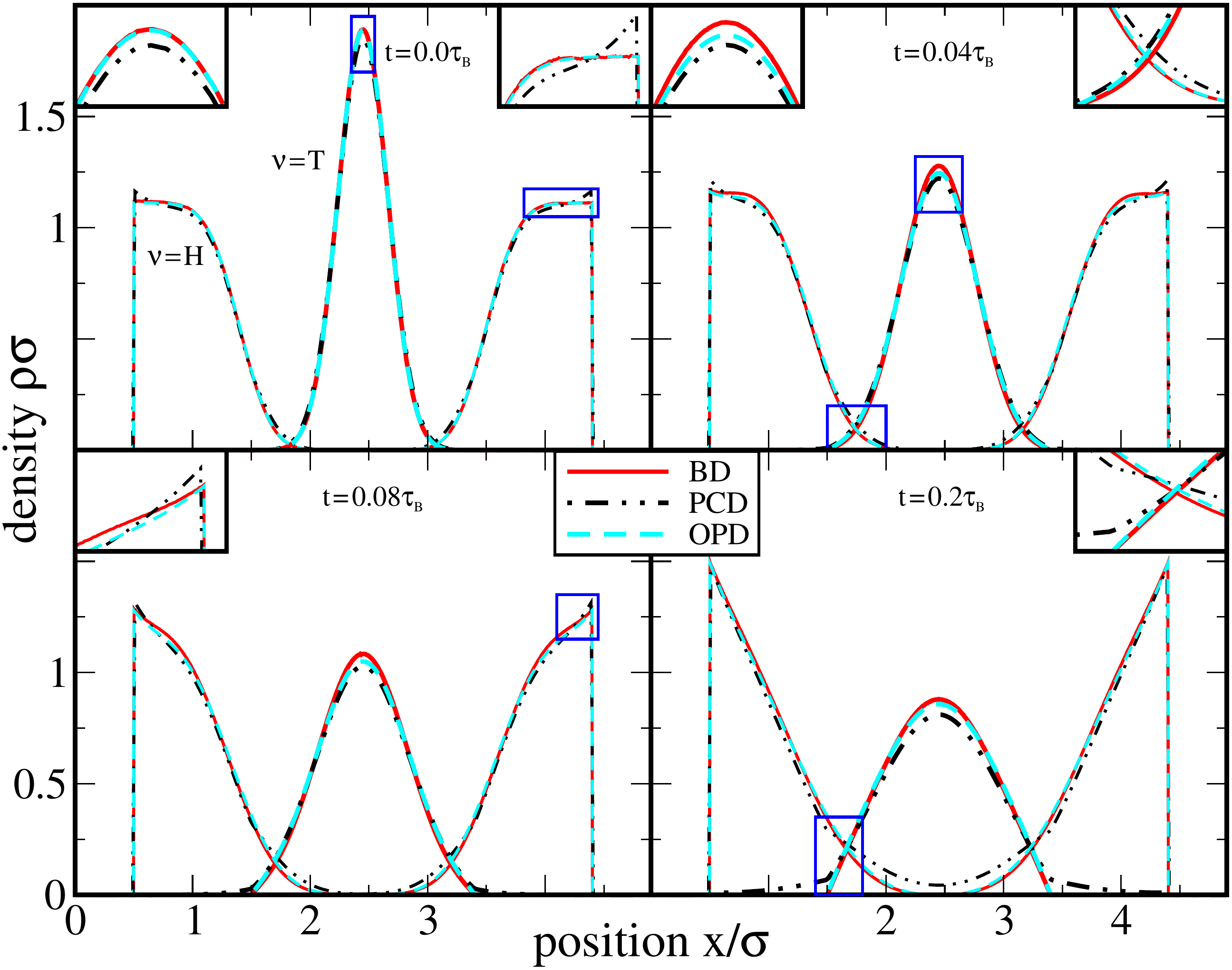} 
\caption{Extension of Fig.~5 from the main manuscript. Shown are the individual density profiles $\rho_N^{(\nu)}(x,t)$ of $N\!=\!3$ hard rods of length $\sigma$ on a line of length $l\!=\!4.9\sigma$ for distinguished initial trapping 
 with $k^{(\TT)}\!=\!20/(\beta\sigma^2)$, $x_\text{h}^{(\TT)}\!=\!l/2$ and $k^{(\LL)}\!=\!k^{(\RR)}\!=\!0$ (species $\nu\!=\!{\LL}$ and $\nu\!=\!{\RR}$ are joined to a single species $\nu\!=\!{\HH}$).
We compare BD and PCD results as in Ref.~\cite{schindlerproject} to OPD as labeled, where each subfigure shows the results for a given time.
The insets show closeups of the boxed regions.
\label{fig_T3figCL}
}
\end{figure*}

\begin{figure*}[t]
\includegraphics[width=0.6\textwidth] {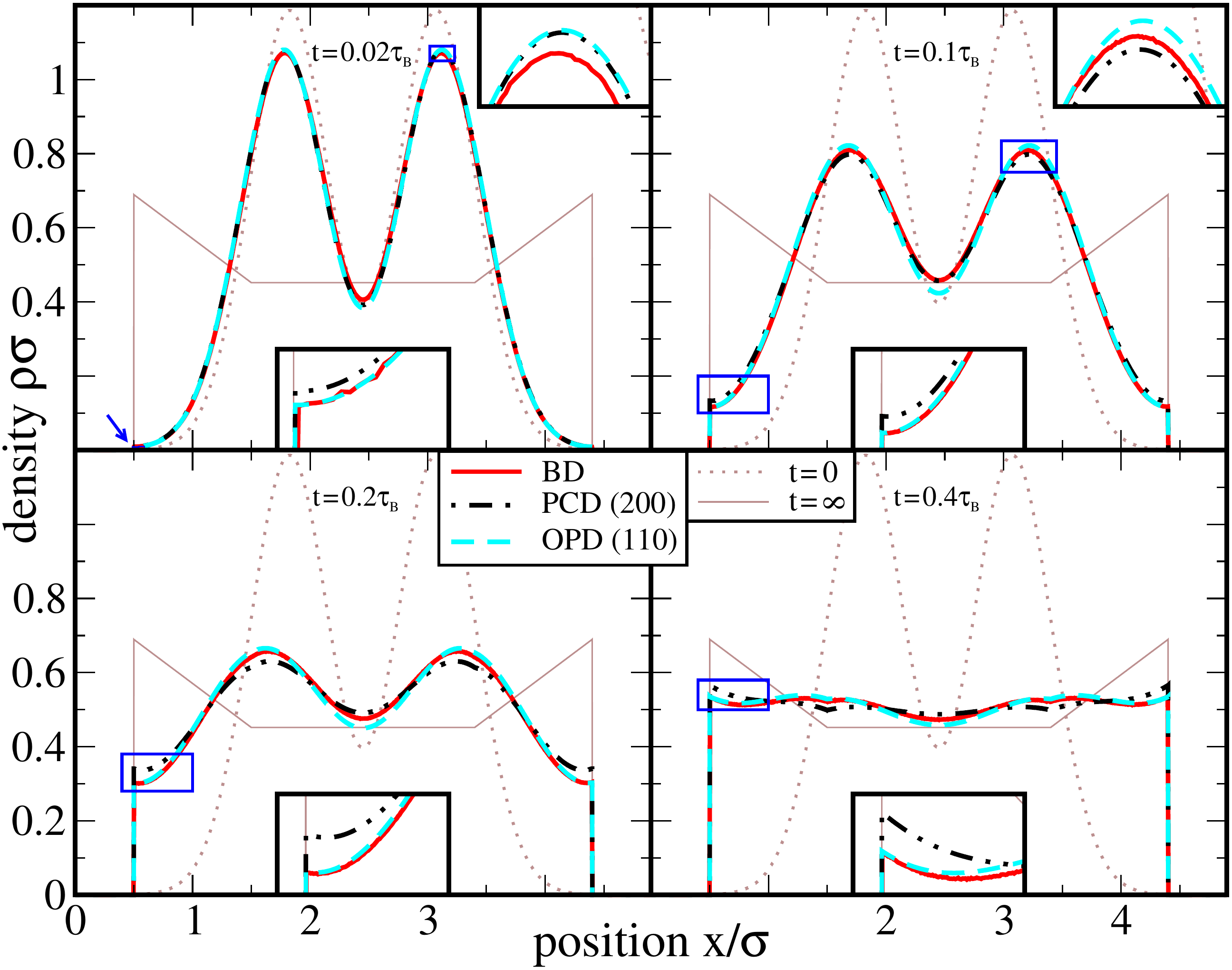} 
\caption{Extension of Fig.~6 from the main manuscript.
Shown are the total density profiles $\rho_N(x,t)$ of $N\!=\!2$ 
 hard rods of length $\sigma$ on a line of length $l\!=\!4.9\sigma$ for common initial trapping 
with $k^{(\nu)}\!=\!5/(\beta\sigma^2)$ and $x_\text{h}^{(\nu)}\!=\!2.45\sigma\!=\!l/2$ using different approaches (as labeled).
The numbers $(N_{\LL}N_{\TT}N_{\RR})$ in brackets indicate the numbers $N_\nu$ of particles in each component $\nu$ of OPD, where PCD is formally equal to OPD if all particles belong to the same species.
The common initial and final profiles are drawn as brown lines (see labels) into each subfigure showing the results for a different finite time.
The insets show closeups of the boxed regions.
\label{fig_T2midCL}
}\end{figure*}
\begin{figure*}[t]
\includegraphics[width=0.6\textwidth] {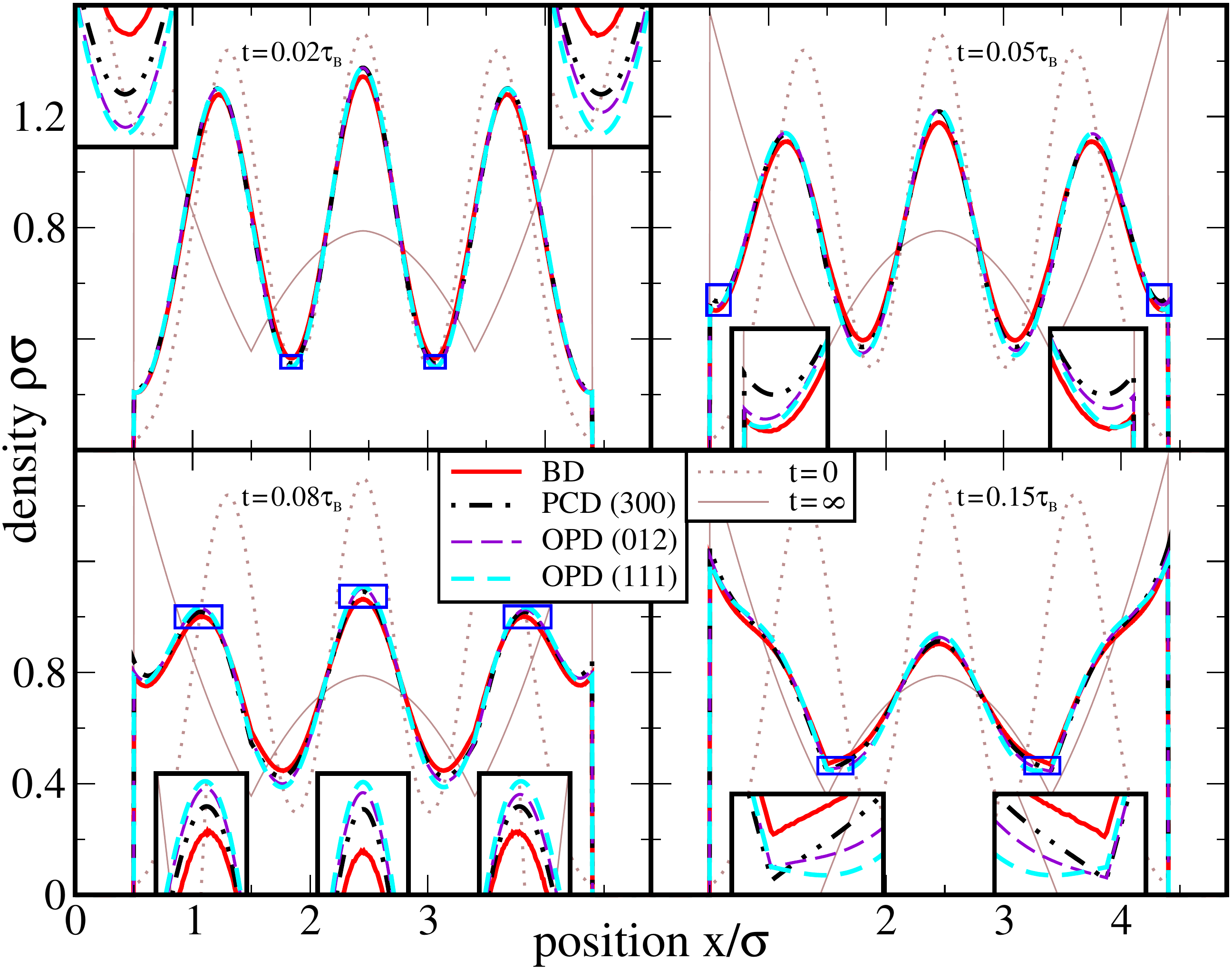} 
\caption{Extension of Fig.~7 from the main manuscript.
Shown are the total density profiles $\rho_N(x,t)$ of $N\!=\!3$ hard rods of length $\sigma$ on a line of length $l\!=\!4.9\sigma$ for common initial trapping 
with $k^{(\nu)}\!=\!5/(\beta\sigma^2)$ and $x_\text{h}^{(\nu)}\!=\!2.45\sigma\!=\!l/2$ using different approaches (as labeled).
The numbers $(N_{\LL}N_{\TT}N_{\RR})$ in brackets indicate the numbers $N_\nu$ of particles in each component $\nu$ of OPD, where PCD is formally equal to OPD if all particles belong to the same species.
One additional set of OPD data (012) is included considering the two particles on the right as members of a single species.
 Joining the left and right particle to a new host species with $\rho_N^{(\HH)}=\rho_N^{(\LL)}+\rho_N^{(\RR)}$ yields the same result as for full OPD with three species.
The common initial and final profiles are drawn as brown lines (see labels) into each subfigure showing the results for a different finite time.
The insets show closeups of the boxed regions. 
\label{fig_T3midCL}
}\end{figure*}

\begin{figure*}[t]
\includegraphics[width=0.6\textwidth] {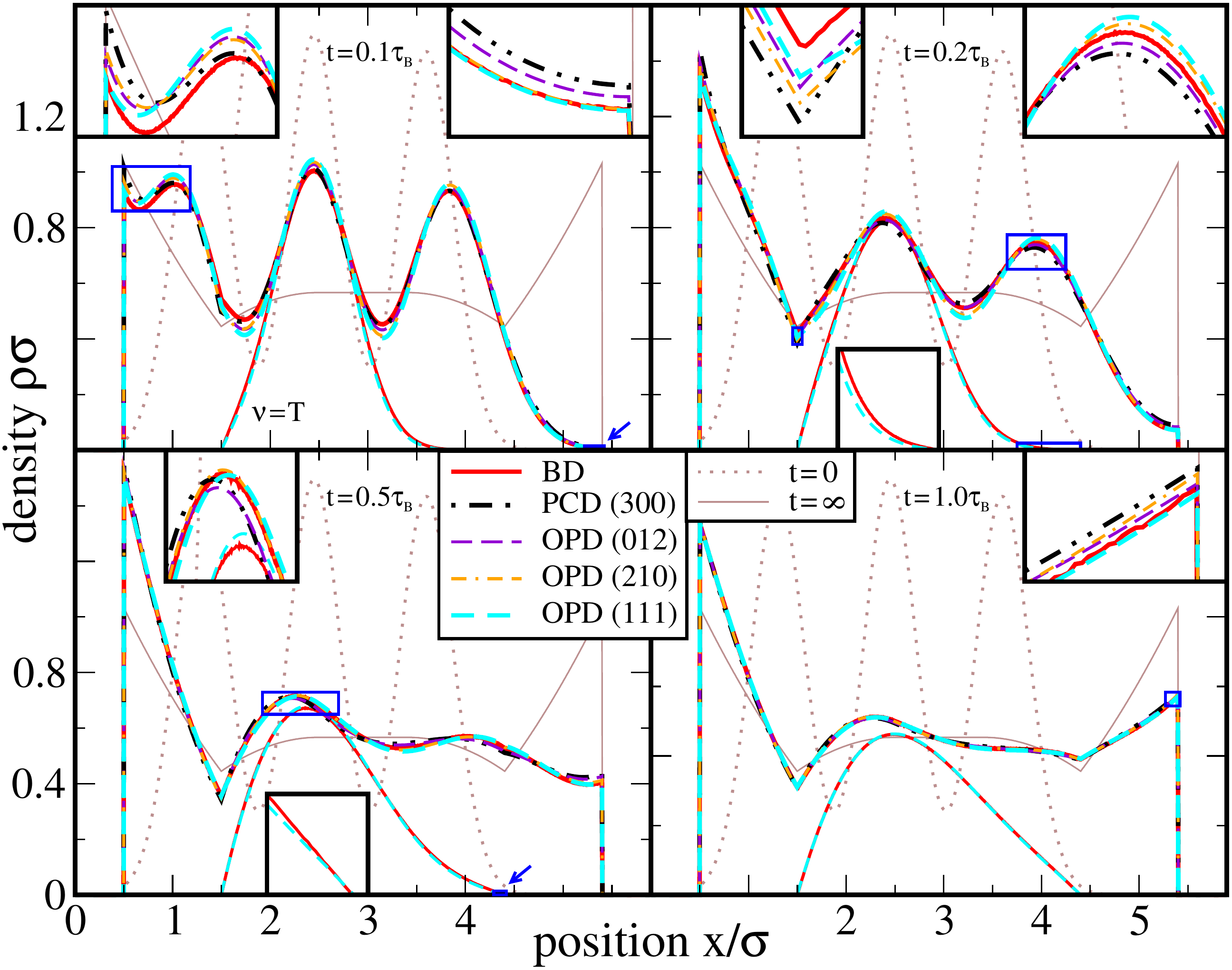} 
\caption{
Extension of Fig.~12 from the main manuscript.
Shown are the total density profiles $\rho_N(x,t)$ of $N\!=\!3$ hard rods of length $\sigma$ on a line of length $l\!=\!5.9\sigma$ for common initial trapping 
with $k^{(\nu)}\!=\!5/(\beta\sigma^2)$ and $x_\text{h}^{(\nu)}\!=\!2.45\sigma\!\neq\!l/2$ using different approaches (as labeled).
The numbers $(N_{\LL}N_{\TT}N_{\RR})$ in brackets indicate the numbers $N_\nu$ of particles in each component $\nu$ of OPD, where PCD is formally equal to OPD if all particles belong to the same species.
Two additional sets of OPD data (012) and (210) are included, each considering two different neighboring particles as members of a single species.
 Joining the left and right particle to a new host species with $\rho_N^{(\HH)}=\rho_N^{(\LL)}+\rho_N^{(\RR)}$ yields the same result as for full OPD with three species.
The common initial and final profiles are drawn as brown lines (see labels) into each subfigure showing the results for a different finite time.
The insets show closeups of the boxed regions. 
\label{fig_T3largeCL}
}\end{figure*}

\begin{figure*}[t]
\includegraphics[width=0.6\textwidth] {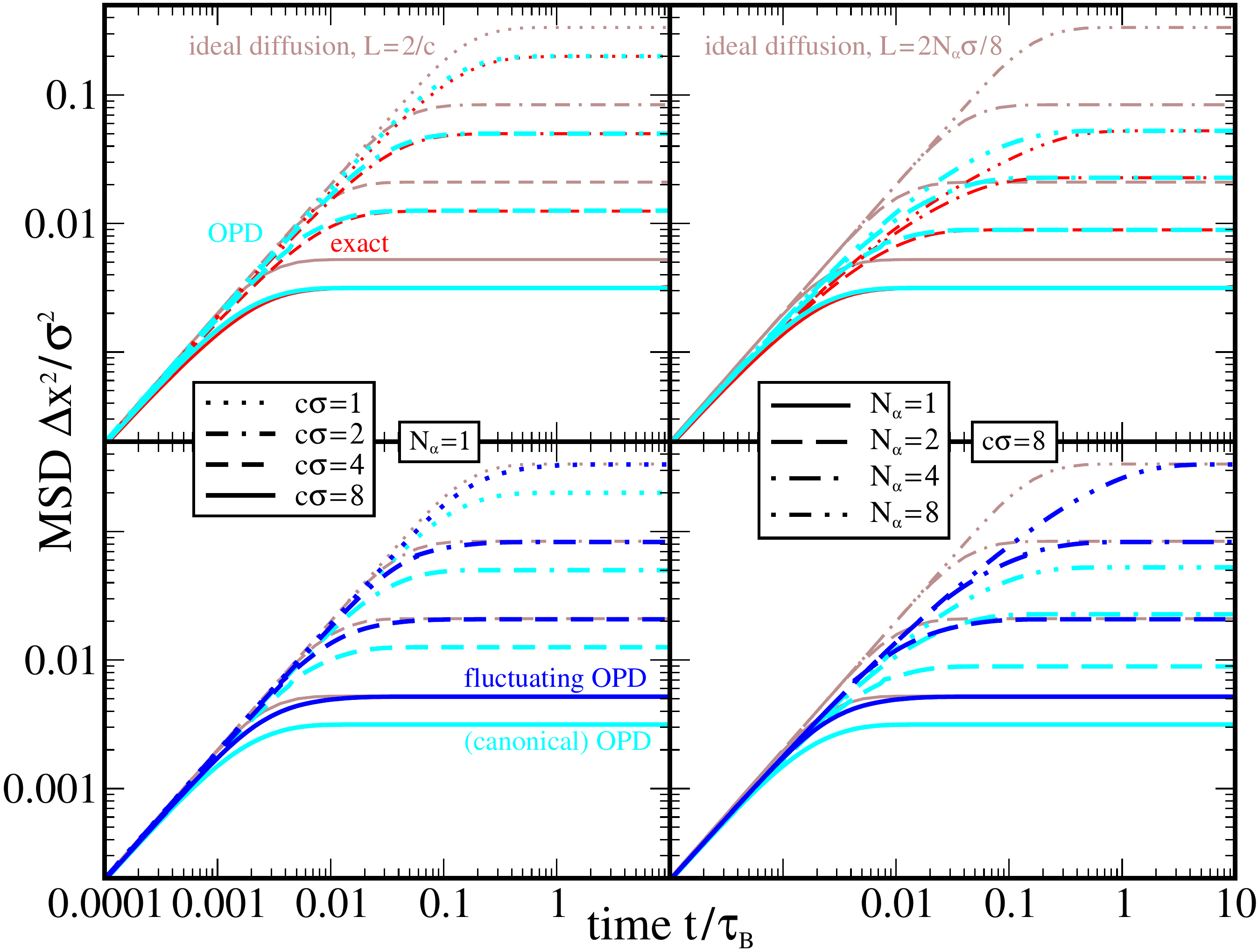} 
\caption{
Extension of Fig.~9 from the main manuscript. 
Shown is the MSD of a tagged point particle in a confined system of length $l$. According to the annotations, we compare the OPD based on the ordered canonical ensemble (thick cyan lines), the fluctuating OPD based on the gcg ensemble (thick blue lines) and exact results (thin red lines). Moreover, we show the MSD of a single (ideal) diffusing particle for $l\!=\!\sigma$ in brown.
The left plots depict the dependence on the concentration $c\!=\!2N_\alpha/l$ of host particles for $N_\alpha\!=\!1$ and the right plots depict the curves for different numbers $N_\alpha$ of host particles at fixed $c\!=\!8/\sigma$ (the solid lines in both cases correspond to the same data). 
The particular value of $l$ and therefore the maximal MSD depends on both $c$ and $N_\alpha$.
\label{fig_MSDboxCL}
}\end{figure*}

\begin{figure*}[t]
\includegraphics[width=0.6\textwidth] {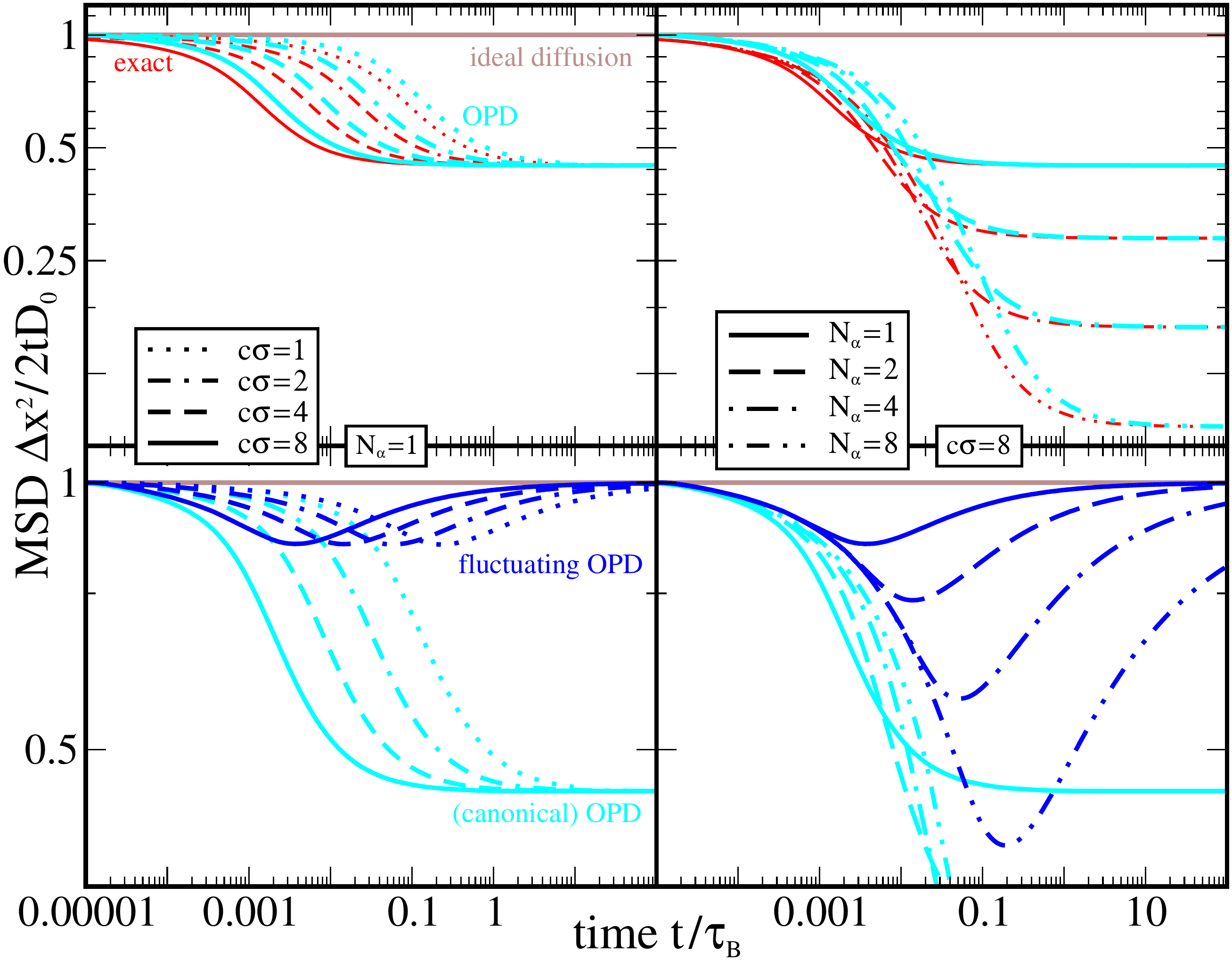} 
\caption{
Extension of Fig.~10 from the main manuscript.
Shown is the MSD of a tagged point particle for an open system with the host particles initially distributed homogeneously in a finite interval of length of $l$.
According to the annotations, we compare the OPD based on the ordered canonical ensemble (thick cyan lines), the fluctuating OPD based on the gcg ensemble (thick blue lines) and exact results (thin red lines).
Notice the different normalization of the vertical axis by the MSD $2D_0t$ of an ideally diffusing free particle (brown horizontal line at unity) to better resolve the differences at intermediate times.
The left plots depict the dependence on the initial concentration $c\!=\!2N_\alpha/l$ of host particles for $N_\alpha\!=\!1$ and the right plots depict the curves for different numbers $N_\alpha$ of host particles at fixed $c\!=\!8/\sigma$ (the solid lines in both cases correspond to the same data). 
The particular value of $l$ depends on both $c$ and $N_\alpha$.
\label{fig_MSDfinCL}
}\end{figure*}

\begin{figure*}[t]
\includegraphics[width=0.6\textwidth] {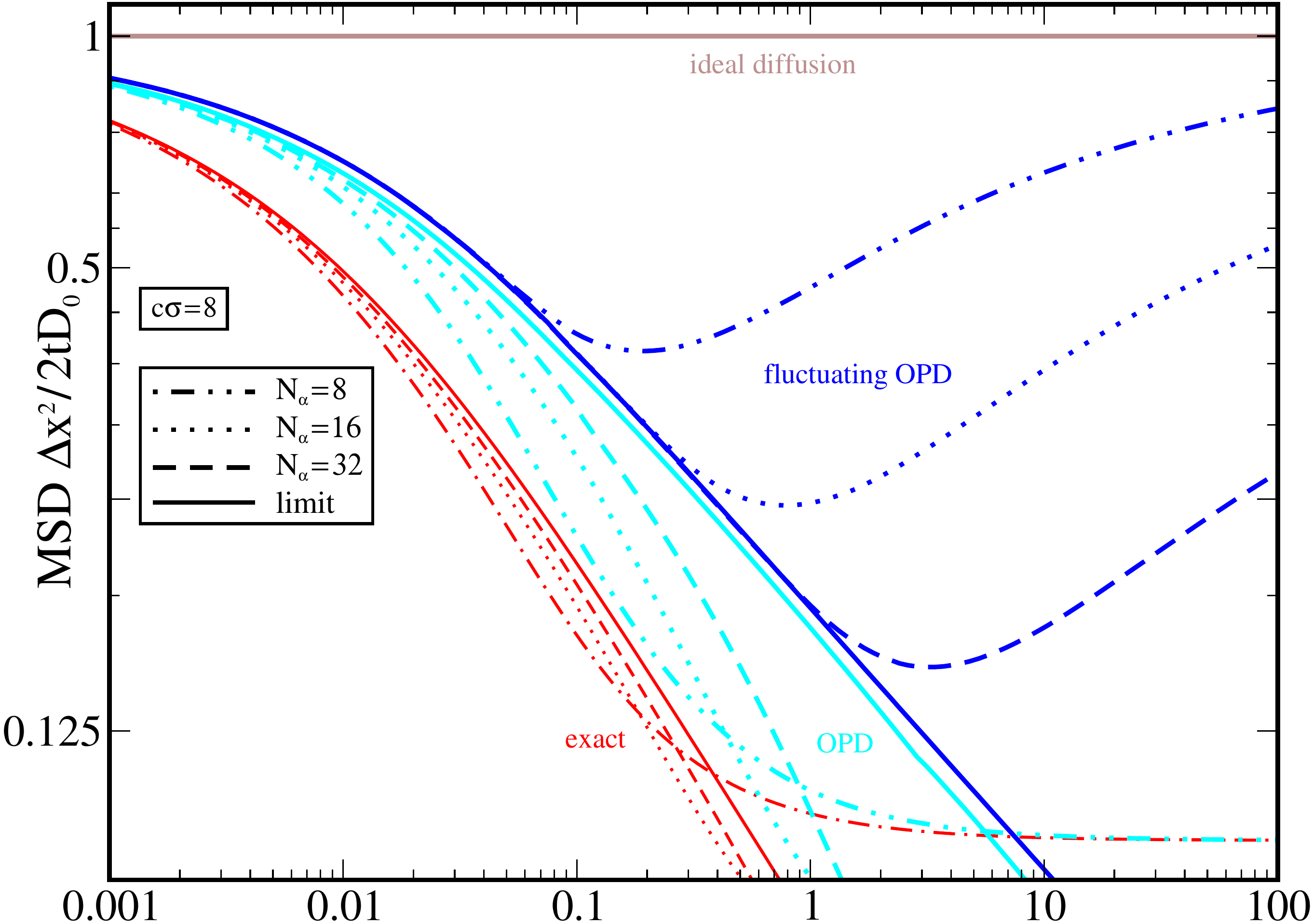} 
\caption{
Connection between Fig.~10 and Fig.~11 from the main manuscript.
Shown is the MSD of a tagged point particle for an open system with the host particles initially distributed homogeneously in a finite interval of length $l\!=\!N_\alpha\sigma/4$ with fixed concentration $c\!=\!8/\sigma$. 
According to the annotations, we compare the OPD based on the ordered canonical ensemble (thick cyan lines), the fluctuating OPD based on the gcg ensemble (thick blue lines) and exact results (thin red lines).
Notice the different normalization of the vertical axis by the MSD $2D_0t$ of an ideally diffusing free particle (brown horizontal line at unity) to better resolve the differences at intermediate times.
The plot depicts the dependence on the numbers $N_\alpha$ of host particles according to the legend. 
The limit $N_\alpha\!\rightarrow\!\infty$ can only be properly evaluated for fluctuating OPD.
However, the approximate result 
for canonical OPD, obtained by subsequently doubling the finite particle number $N_\alpha$ (starting from $N_\alpha\!=\!16$) and the total system size,
provides a strong evidence for the equivalence of both OPD approaches in this limit. 
\label{fig_MSDlimCL}
}\end{figure*}


\begin{thebibliography}{10}
\bibitem{noteGC} Note that the grand-canonical intrinsic free energy functional $\mathcal{F}[\{\rho^{(\nu)}(x)\}]$ expressed in a form analogous to Eq.~\eqref{eq_Fcan} contains additional terms including the chemical potentials $\mu_\nu$, which are linear in the densities and thus vanish after taking the spatial derivative of the functional derivative.
\end{thebibliography}
\end{document}